\documentclass[conference]{IEEEtran}
\usepackage[utf8]{inputenc}

\usepackage{tikz}
\usepackage{pgfplots}
\usepackage[keeplastbox]{flushend}
\usepackage{aircraftshapes}
\usetikzlibrary{angles,quotes}
\pgfplotsset{compat=newest}
\pgfplotsset{every axis legend/.append style={%
		cells={anchor=west}}
}
\pgfplotsset{every y tick label/.append style={font=\footnotesize}}
\pgfplotsset{every x tick label/.append style={font=\footnotesize}}
\pgfplotsset{every axis x label/.append style={font=\footnotesize}}
\pgfplotsset{every axis y label/.append style={font=\footnotesize}}
\pgfplotsset{every axis legend/.append style={font=\footnotesize}}
\pgfplotsset{every axis title/.append style={font=\footnotesize}}

\usepgfplotslibrary{polar}
\usetikzlibrary{arrows}
\tikzset{>=stealth'}
\usepgfplotslibrary{groupplots}
\usepackage[per-mode=symbol,detect-all]{siunitx}
\DeclareSIUnit\feet{ft}

\DeclareGraphicsExtensions{.pdf, .jpg, .png}
\usepackage{amsmath}
\usepackage[noend]{algorithmic}
\usepackage{algorithm}
\usepackage{array}
\usepackage[caption=false,font=normalsize,labelfont=sf,textfont=sf]{subfig}
\usepackage{booktabs}
\usepackage[capitalize]{cleveref}
\usepackage{todonotes}
\newcommand{\argmax}{\operatornamewithlimits{arg\,max}}

\usepackage[backend=biber,style=ieee,,maxcitenames=1,mincitenames=1]{biblatex}
\makeatletter
\def\blx@maxline{77}
\makeatother

\AtEveryBibitem{
	\ifentrytype{inproceedings}{
		\clearlist{address}
		\clearlist{publisher}
		\clearname{editor}
		\clearlist{organization}
		\clearfield{url}  
		\clearfield{url}  
		\clearfield{doi}  
		\clearfield{pages}  
		\clearlist{location}}{}}
\AtEveryBibitem{
	\clearfield{doi}
	\clearfield{issn}
	\clearfield{month}  
}
\AtEveryBibitem{
	\ifentrytype{article}{
		\clearlist{address}
		\clearlist{publisher}
		\clearname{editor}
		\clearlist{organization}
		\clearfield{url}  
		\clearfield{doi}  
		\clearlist{location}}{}}

\title{Guaranteeing Safety for Neural Network-Based \\ Aircraft Collision Avoidance Systems}

\author{\IEEEauthorblockN{Kyle D. Julian and
Mykel J. Kochenderfer}
\IEEEauthorblockA{Department of Aeronautics and Astronautics, 
Stanford University, Stanford, CA, 94305}}

\begin{document}

\maketitle

\begin{abstract}
	The decision logic for the ACAS X family of aircraft collision avoidance systems is represented as a large numeric table. Due to storage constraints of certified avionics hardware, neural networks have been suggested as a way to significantly compress the data while still preserving performance in terms of safety. However, neural networks are complex continuous functions with outputs that are difficult to predict. Because simulations evaluate only a finite number of encounters, simulations are not sufficient to guarantee that the neural network will perform correctly in all possible situations. We propose a method to provide safety guarantees when using a neural network collision avoidance system. The neural network outputs are bounded using neural network verification tools like Reluplex and Reluval, and a reachability method determines all possible ways aircraft encounters will resolve using neural network advisories and assuming bounded aircraft dynamics. Experiments with systems inspired by ACAS X show that neural networks giving either horizontal or vertical maneuvers can be proven safe. We explore how relaxing the bounds on aircraft dynamics can lead to potentially unsafe encounters and demonstrate how neural network controllers can be modified to guarantee safety through online costs or lowering alerting cost. The reachability method is flexible and can incorporate uncertainties such as pilot delay and sensor error. These results suggest a method for certifying neural network collision avoidance systems for use in real aircraft.
\end{abstract}

%\todo[inline]{KJ: Is the last sentence too strong?}
%\todo[inline]{KJ: How important is it to show operation and safety metrics? I'm running out of room in this paper and don't know if they add much of value besides mentioning that retraining a network with a lower alerting cost may help verification, but this will result in more alerts}
%  Operational and safety metrics compiled through simulation demonstrate the trade-off between guaranteeing safety in all possible cases and minimizing the number of advisories given.

\section{Introduction}

The ACAS X family of next-generation collision avoidance systems represents optimized solutions to Markov decision processes (MDPs) using large numeric lookup tables~\cite{kochenderfer2011robust,Kochenderfer2015chapter10}. The version for large manned aircraft, ACAS Xa, issues vertical advisories to pilots and has become an international standard~\cite{Kochenderfer2015chapter10}. Another version for unmanned aircraft, ACAS Xu, issues turn rate advisories to remote pilots to avoid near midair collisions (NMACs), and demonstrations have been flown with NASA's Ikhana aircraft~\cite{ACAS-XuTests}.

The large size of the lookup table can be challenging for current avionic systems with limited memory. \citeauthor{julian2018deep} demonstrate that a neural network approximation of the table can reduce representation size by a factor of 1000 without increasing online computation time or diminishing operational performance in simulation~\cite{julian2018deep}. However, simulations check only a finite number of encounter geometries and do not guarantee that all possible encounters will be safely resolved. While neural networks are efficient global function approximators, their output is difficult to guarantee, which can prohibit their use in safety-critical systems.

Recent work on neural network verification has produced many tools and techniques for providing guarantees of neural network outputs given a bounded set of inputs. One tool, Reluplex, extends the simplex algorithm to incorporate rectified linear activation functions to search for a set of inputs that satisfies all equations and constraints~\cite{katz2017reluplex}. Another tool, Reluval, uses symbolic bound propagation and a splitting function to verify network properties~\cite{wang2018formal}. Many other tools exist that use ideas from optimization, reachability, and search to verify properties~\cite{liu2019algorithms}.

These tools verify simple input-output properties for neural networks. \citeauthor{katz2017reluplex} test prototype neural networks trained on an early version of the ACAS Xu logic table to show that intuitive properties hold, such as a strong right turn is advised when the intruder approaches from the left~\cite{katz2017reluplex}. However, these properties are only sanity checks that ought to hold, but they do not guarantee that safety will be maintained over time.

To guarantee safety, verification must include the system dynamics to understand how the closed-loop system evolves in time. \citeauthor{julian2019verifying} use the system dynamics to derive input-output network properties that guarantee closed-loop safety~\cite{julian2019verifying}. \citeauthor{akintunde2018reachability} use mixed-integer linear programming to prove closed-loop properties of neural network agents acting in neural network approximated environments~\cite{akintunde2018reachability}. \citeauthor{ivanov2018verisig} use ideas from hybrid systems for verification of closed-loop properties of sigmoid-activated neural networks~\cite{ivanov2018verisig}. \citeauthor{xiang2018reachability} use a reachability method to over-approximate a neural network controller's output along with reachability tools to bound state variable trajectories~\cite{xiang2018reachability}. \citeauthor{julian2019reachability} use reachability combined with neural network verification tools to prove closed-loop properties~\cite{julian2019reachability}.

This paper presents a reachability method that over-approximates the neural network and system dynamics to check whether an NMAC is reachable. If the over-approximated system cannot reach an NMAC, then the real system is guaranteed safe assuming the dynamical model constraints hold. In addition, because some of the ACAS X collision avoidance systems are still under development and not publicly available, this paper presents two open-source collision avoidance systems, VerticalCAS and HorizontalCAS, that are notional examples inspired by early versions ACAS Xa and ACAS Xu. These examples formulate collision avoidance systems as an MDP and train neural network representations. The systems and supporting code are available at \url{https://github.com/sisl/VerticalCAS} and \url{https://github.com/sisl/HorizontalCAS}. Although this paper studies the verification of notional systems, the methods and lessons learned can still be applied to real collision avoidance systems or other neural network controllers.

The reachability method verifies safety for both collision avoidance systems. The dynamic constraints are relaxed until safety can no longer be guaranteed to determine the safe range of aircraft accelerations and turn rates. %\todo{this sentence is a bit unclear}. 
The reachability method is flexible, and this work explores how uncertainties such as pilot delay and sensor error can be incorporated into the reachability framework and proven safe. Additionally, reachability can determine the region where the decisions do not affect safety, which could be used to speed the verification process by focusing on only safety-critical regions of the state space. These results demonstrate how neural network controllers in real safety-critical systems can be proven safe, which could play a role in their verification and incorporation into real systems.

This paper describes the general collision avoidance table generation and neural network approximation methods in \Cref{sec:cas}. \Cref{sec:VerticalCAS} and \Cref{sec:HorizontalCAS} describe the VerticalCAS and HorizontalCAS collision avoidance systems respectively. \Cref{sec:Verif} presents the reachability method for safety verification, while \Cref{sec:VertResults} and \Cref{sec:HorResults} describe the reachability results using VerticalCAS and HorizontalCAS, and \Cref{sec:conc} concludes.

%The reachability method is a formal method that checks if an NMAC ever occurs.
%Because the MDP used to generate the logic table uses a probabilistic model that balances the cost of alerting with the probability of an NMAC, formally guaranteeing no NMACS could ever occur must use a more constrained dynamic model, otherwise reachability will show that 

 %ACAS Xa uses a lookup table of state-action values, but the table is under current development and cannot be publicly shared.
 
 %Neural network verification tools such as Reluplex~\cite{katz2017reluplex} and Reluval~\cite{wang2018formal} can be used to compute if an advisory is given at any point within a given region of the input space. \citeauthor{katz2017reluplex} prove simple input-output properties for neural networks trained on a prototype of ACAS Xu~\cite{katz2017reluplex}. Examples of these properties are the networks always advises COC when the intruder is  far away, or the networks always advises a strong right turn when the intruder is approaching from the left. However, these properties only describe sanity checks that should intuitively hold, but they do not guarantee that safety will be maintained when using the neural network system.

\section{Collision Avoidance Systems}\label{sec:cas}
One method for generating the logic for collision avoidance systems uses dynamic programming to compute scores for all advisories in all discrete encounter states. The table grows exponentially with the number of state variables, making it challenging to store in certified avionic hardware. A neural network can be trained to approximate the logic table and significantly reduce required storage space~\cite{julian2018deep}.

\subsection{Table Generation}
The collision avoidance score table is generated by framing the collision avoidance problem as a Markov decision process (MDP)~\cite{bellman1952theory, Kochenderfer2015chapter4}. An MDP is composed of states $s \in \mathcal{S}$, actions $a \in \mathcal{A}$, a reward function $R(s,a)$, and state transition probabilities $T(s' \mid s,a)$ that define the probability of reaching state $s'$ by taking action $a$ in state $s$. The solution to an MDP is a policy $\pi(s)$ that  maps states to actions in order to maximize the accumulation of reward over time. The policy $\pi(s)$ can be computed through dynamic programming by defining state-action values $Q(s,a)$, initially all zeros, and recursively updating $Q$ using the Bellman equation,
\begin{equation} \label{eq:bell}
    Q(s,a) \gets R(s,a) + \gamma\sum_{s' \in \mathcal{S}} T(s' \mid s,a) \max_{a' \in \mathcal{A}} Q(s',a')\text{,}
\end{equation}
where $\gamma$ is a discount factor to ensure convergence. After computing $Q(s,a)$, we may determine the policy $\pi$ from 
\begin{equation}
    \pi(s) = \argmax_{a \in \mathcal{A}} Q(s,a).
\end{equation}

The state-action values $Q(s,a)$ form the logic table to be compressed with a neural network. Although the network could approximate $\pi(s)$ directly instead of all state-action values, systems like ACAS X use the values to compute the best action in multi-intruder encounters, so this work approximates the state-action values rather than just the policy~\cite{kochenderfer2011robust, julian2018deep}.

\subsection{Neural Network Approximation}
After computing the state-action value table $Q(s,a)$, a neural network representation denoted $\tilde{Q}(s,a)$ is used to approximate the table. Previous work shows that a neural network representation maintains accuracy while reducing representation size and outperforms other compression methods such as decision trees or symmetry analysis~\cite{julian2016policy,julian2018deep}.

Neural networks are composed of layers, where the input layer is the encounter geometry, the output layer contains the action scores, and the middle layers are called hidden layers. Each hidden layer is computed as an affine transformation of the previous layer before applying a nonlinear activation function. This work uses networks with rectified linear unit activations (ReLU), defined as $\text{relu}(x)=\max(0,x)$~\cite{ReLU}. Defining $x_0$ as the input to the network with $L$ hidden layers, the hidden layers can be computed as 
\begin{equation}\label{eq:network}
    x_{i+1} = \text{relu} \left( W_i x_i+b_i \right) \ \forall  i\in\{0,1,\dots,L-1\}\text{,}
\end{equation}
where $W_i$ and $b_i$ are trainable network parameters. The network output $\tilde{Q}(s,\cdot)$ is computed without any activation function, so $\tilde{Q}(s,\cdot)=W_Lx_L + b_L$.

The network parameters are initialized randomly and updated with gradient descent methods to minimize a loss function representing the error of the network. Typical regression problems use a mean squared error loss function; however, approximating $Q(s,a)$ should also strive to maintain $\argmax_{a \in \mathcal{A}} Q(s,a)$ so that $\pi(s)$ is accurately approximated as well. Nominal mean squared error weights both under- and over-approximation errors equally, which could result in changes to $\pi(s)$ if the value of the best action is under-valued while another action is over-valued. 

This work uses an asymmetric mean squared error function, which heavily penalizes under-valuing the best action or over-valuing the other actions~\cite{julian2018deep}. In addition, a linear term is added to the loss function for under-valuing the best action or over-valuing the other actions, which strengthens the gradient when error is small. The loss function $\ell(s,a)$ can be written as function of the prediction error $e(s,a)=\tilde{Q}(s,a)-Q(s,a)$ using
\begin{equation}
    \ell(s,a) = 
    \begin{cases}
    e(s,a)^2, &\text{if}\ a\neq \pi(s) \ \land \\ & \quad e(s,a)<0 \\
    e(s,a)^2, &\text{if}\ a=\pi(s) \ \land \\ & \quad e(s,a)\ge0 \\
    c \left( e(s,a)^2+ \vert e(s,a) \vert \right), &\text{if}\ a\neq\pi(s) \ \land\\ & \quad e(s,a)\ge0 \\
    c(N-1) \left( e(s,a)^2+ \vert e(s,a) \vert \right), &\text{otherwise}
    \end{cases}
\end{equation}
where $c$ is a scaling constant and $N$ is the number of network outputs. Increasing the scaling factor $c$ penalizes the network more if errors $e(s,a)$ could change the policy, and the $N-1$ term balances the fact that there are $N-1$ actions for $a\neq\pi(s)$ but only one action for $a=\pi(s)$. In this work, $c=40$.

For each epoch in supervised learning, state-action values from the table are randomly split into smaller batches of 512, and each batch is used to compute an average $\ell(s,a)$. Then, the neural network parameters are updated through the adaptive gradient descent method Adam. The neural networks are trained for multiple epochs until improvement stagnates.

%%%%%%%%%%%%%%%
% VerticalCAS %
%%%%%%%%%%%%%%%
\section{VerticalCAS}\label{sec:VerticalCAS}
%VerticalCAS is a notional example inspired by a prototype of the ACAS Xa collision avoidance system~\cite{julian2019verifying}. VerticalCAS closely follows the prototype ACAS Xa description provided by \citeauthor{Kochenderfer2015chapter10} with a few simplifications~\cite{Kochenderfer2015chapter10}. Supporting code for VerticalCAS can be found at \url{https://github.com/sisl/VerticalCAS}.

VerticalCAS issues vertical rate advisories to an aircraft, called the ownship, to avoid near midair collisions (NMACs) with another aircraft, called an intruder. NMACs are defined as separation less than 100 ft vertically and 500 ft horizontally. 

\begin{table}
    \centering
    \caption{VerticalCAS state variables}
    \begin{tabular}{lccc}  
	\toprule
	Variable  & Description & Values & Num \\
	\midrule
	$h$ (\si{\feet}) & Relative intruder altitude & $[-8000,8000]$ & 65 \\
	$\dot{h}_\text{own}$ (\si{\feet\per\minute}) & Ownship vertical rate & $[-6000,6000]$ & 39 \\
	$\dot{h}_\text{int}$ (\si{\feet\per\minute}) & Intruder vertical rate & $[-6000,6000]$ & 39 \\
	$\tau$ (\si{\second}) & Time to loss of hor. separation & $[0,40]$ & 41\\
	$s_\text{adv}$ & Previous advisory & See \Cref{tab:VertAdv} & 9 \\
	\bottomrule
\end{tabular}\label{tab:VertStates}
\end{table}

\begin{figure}
	\centering
	\begin{tikzpicture}
\node [aircraft side,fill=black,minimum width=1cm,rotate=10] (own) at (0,0) {};
\coordinate[label=below:Ownship] (ownText) at (0,-0.3);
\node [aircraft side,draw=black,fill=red,minimum width=1cm,xscale=-1] (int) at (4,1.3) {}; 
\coordinate[label=right:Intruder] (intText) at (4.65,1.37);

\draw (0.45,0.09) edge[->] node[midway,below] {$\tau$} (3.58,0.09);
\draw [dashed] (3.6,1.45) -- (3.6,-0.35);
\draw [dashed] (3.6,0.09) -- (4.1,0.09);
\draw (0.0,0.12) edge[->] node[midway,right] {$\dot{h}_\text{own}$} ++(0.0,1.0) ;
\draw (3.85,1.42) edge[->] node[midway,right] {$\dot{h}_\text{int}$} ++(0.0,1.0) ;
\draw (3.85,0.12) edge[->] node[midway,right] {$h$} ++(0.0,1.16) ;
\end{tikzpicture}
	\caption{Aircraft encounter geometry for VerticalCAS}\label{fig:VerticalCAS}
\end{figure}
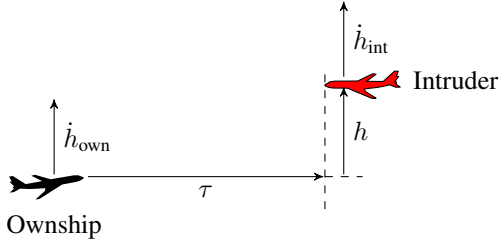

The state space is composed of five dimensions, as described in \Cref{tab:VertStates}. The variables $h$, $\dot{h}_\text{own}$, and $\dot{h}_\text{int}$ describe the vertical encounter geometry. The variable $\tau$ captures the horizontal geometry and acts as a countdown to the time when horizontal separation is lost, at which point the aircraft must be separated vertically to avoid an NMAC. The variable $s_\text{adv}$ represents the previous advisory given, allowing the system to choose new alerts consistently. The encounter geometry is shown in \Cref{fig:VerticalCAS}.

\begin{table}
    \centering
    \caption{VerticalCAS advisories}\label{tab:VertAdv}
    \setlength\tabcolsep{3.2pt}
\begin{tabular}{lcc}  
	\toprule
	Advisory  & Description & $\ddot{h}_\text{own}$ range $(\si{\feet\per\second\squared})$ \\
	\midrule
	COC     &  Clear of Conflict  & $\left[-10.7,10.7\right]$    \\
	DNC     &  Do Not Climb    & $\left[-16.1,-8.33\right]$  \\
	DND     &  Do Not Descend  & $\left[8.33,16.1\right]$   \\
	DES1500 & Descend $\ge \SI{1500}{\feet\per\minute}$ & $\left[-16.2,-8.33\right]$  \\ 
	CL1500 & Climb $\ge \SI{1500}{\feet\per\minute}$  & $\left[8.33,16.2\right]$ \\ 
	SDES1500 & Strengthen Descent to $\ge \SI{1500}{\feet\per\minute}$  & $\left[-16.2,-10.7\right]$ \\ 
	SCL1500 & Strengthen Climb to $\ge \SI{1500}{\feet\per\minute}$ & $\left[10.7,16.2\right]$ \\ 
	SDES2500 & Strengthen Descent to $\ge \SI{2500}{\feet\per\minute}$& $\left[-16.2,-10.7\right]$ \\ 
	SCL2500 & Strengthen Climb to $\ge \SI{2500}{\feet\per\minute}$  & $\left[10.7,16.2\right]$ \\ 
	
	\bottomrule
\end{tabular}
\end{table}

The action space, described in  \cref{tab:VertAdv}, consists of nine pilot advisories given at a frequency of once per second that dictate the ownship's acceleration. While COC allows the ownship to choose any acceleration, the other advisories assume that the pilot accelerates at an allowed acceleration until complying with the target vertical rate, at which point the pilot stops accelerating. If the aircraft already complies with the advisory, then the aircraft is assumed to maintain the current vertical rate.

The dynamic model can be written as
\begin{equation}
    \begin{bmatrix}
h \\
\dot{h}_\text{own} \\
\dot{h}_\text{int} \\
\tau \\
s_{\text{adv}} \\
\end{bmatrix}
\gets
\begin{bmatrix}
h + \dot{h}_\text{int} + 0.5\ddot{h}_\text{int} - \dot{h}_\text{own} - 0.5\ddot{h}_\text{own} \\
\dot{h}_\text{own} + \ddot{h}_\text{own}  \\
\dot{h}_\text{int} + \ddot{h}_\text{int}  \\
\max(0,\tau-1) \\
s'_{\text{adv}}
\end{bmatrix}\text{,}\label{eq:VertCAS_dynamics}
\end{equation}
where the ownship acceleration $\ddot{h}_\text{own}$ is constrained by $s_\text{adv}$. The intruder is assumed to accelerate in the range $\left[-g/8,\ g/8\right]$. To compute $T(s' \mid s,a)$, a simple distribution is used where the mean acceleration has probability 0.5 while the extremes have probability 0.25. Therefore, the distribution $T(s' \mid s,a)$ will have nine non-zero entries for the different combinations of ownship and intruder accelerations.

The reward function encourages the system to issue advisories that maintain safety while limiting unnecessary alerts and undesired behavior, which includes reversing the advisory direction, strengthening or weakening the advisory, and crossing advisories that make the ownship pass through the intruder's altitude.

Each state dimension of the VerticalCAS state space is discretized into a number of discrete values as specified in \Cref{tab:VertStates}, resulting in 36.4 million states. The MDP is solved in a few minutes using Gauss-Seidel value iteration, resulting in a large state-action score table~\cite{Kochenderfer2015chapter4}. As described in \Cref{sec:cas}, the score table can be approximated by neural networks. This work trains a separate neural network for each $s_\text{adv}$, resulting in nine networks. In addition, the reachability analysis discussed in \Cref{sec:VertResults} studies properties for intruder aircraft that maintain level flight, so the intruder vertical rate input was removed to simplify the networks, though future work could incorporate intruder vertical rate. The networks have three inputs, one for each remaining state variable. Each network has five hidden layers of length 25 and nine outputs, one for each advisory. Nine smaller networks are trained instead of a single large network to reduce run-time required to evaluate each network~\cite{julian2018deep}. Each network was trained for 3000 epochs in 30 minutes using Tensorflow, resulting in nine neural networks that reduce required storage from 1.22GB to 103KB while maintaining the correct advisory 94.9\% of the time.

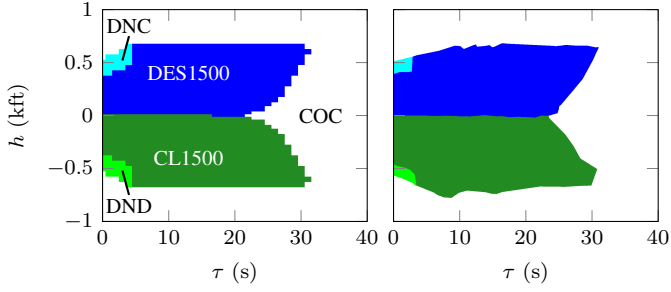
\begin{figure}
	\centering
	\begin{tikzpicture}[]
\begin{groupplot}[height={4.4cm}, width={5.1cm},group style={horizontal sep=0.35cm, group size=2 by 1}]
\nextgroupplot [xlabel = {$\tau\ (\si{\second})$},ylabel = {$h\ (\si{\kilo\feet})$}, xmin = {0.0}, xmax = {40.0}, ymax = {1.0}, ymin = {-1.0}, enlargelimits = false, axis on top]\addplot [point meta min=1, point meta max=9] graphics [xmin=0.0, xmax=40.0, ymin=-1.0, ymax=1.0] {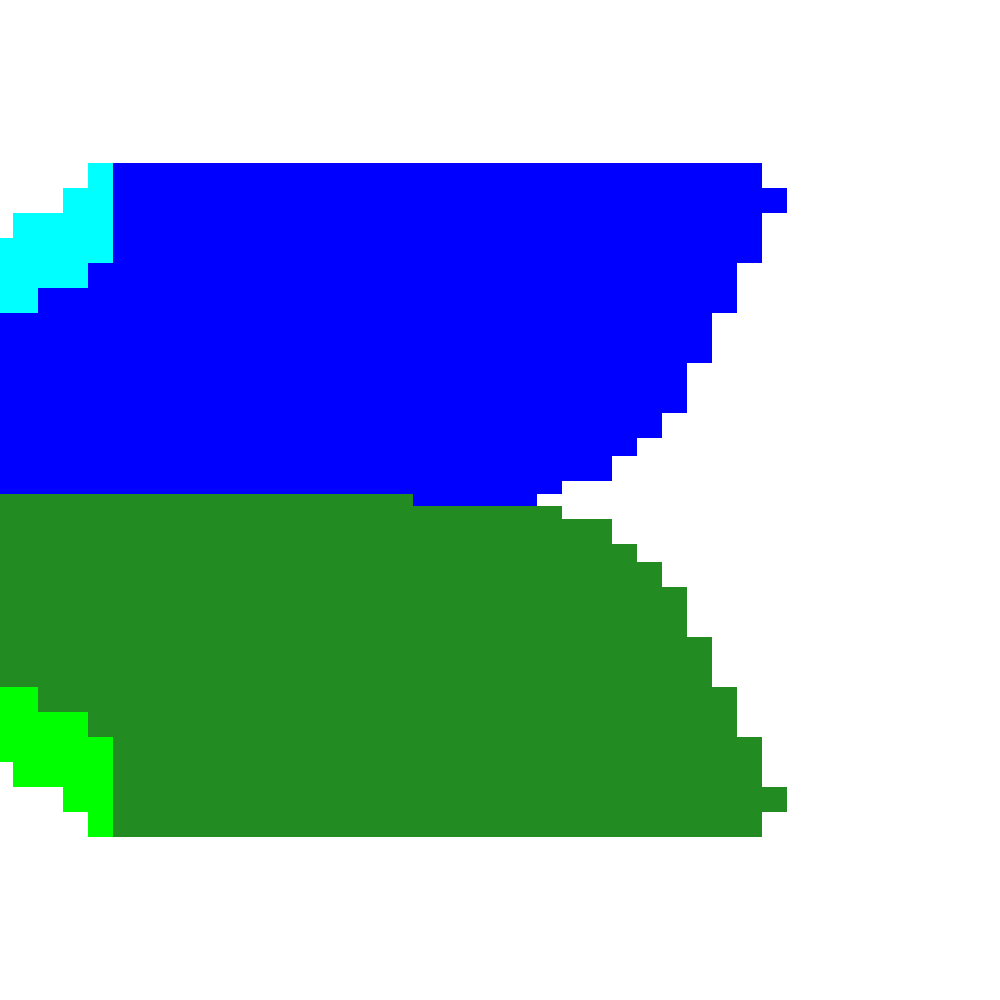};

\node[text=black] at (33,0)  {\footnotesize COC};
\node[text=white] at (13,-.400)  {\footnotesize CL1500};
\node[text=white] at (13,.400)  {\footnotesize DES1500};
\node[text=black] at (4,.850)  {\footnotesize DNC};
\node[text=black] at (4,-.85)  {\footnotesize DND};

\draw[thick] (4,0.75) -- (3,0.52);
\draw[thick] (4,-0.75) -- (3,-0.52);

%\node[aircraft side,fill=black,draw=white,minimum width=1cm,rotate=0,scale = 0.8] at (axis cs:3.0, 0.0) {};
%\node[aircraft side,fill=red,draw=black,minimum width=1cm,rotate=0,scale = 0.8,xscale=-1] at (axis cs:35.0, 0.850) {};

\nextgroupplot [scaled y ticks=false, yticklabels={,,}, xmin = {0.0}, xmax = {40.0}, ymax = {1.0}, xlabel = {$\tau\ (\si{\second})$}, ymin = {-1.0}, enlargelimits = false, axis on top]\addplot [point meta min=1, point meta max=9] graphics [xmin=0.0, xmax=40.0, ymin=-1.0, ymax=1.0] {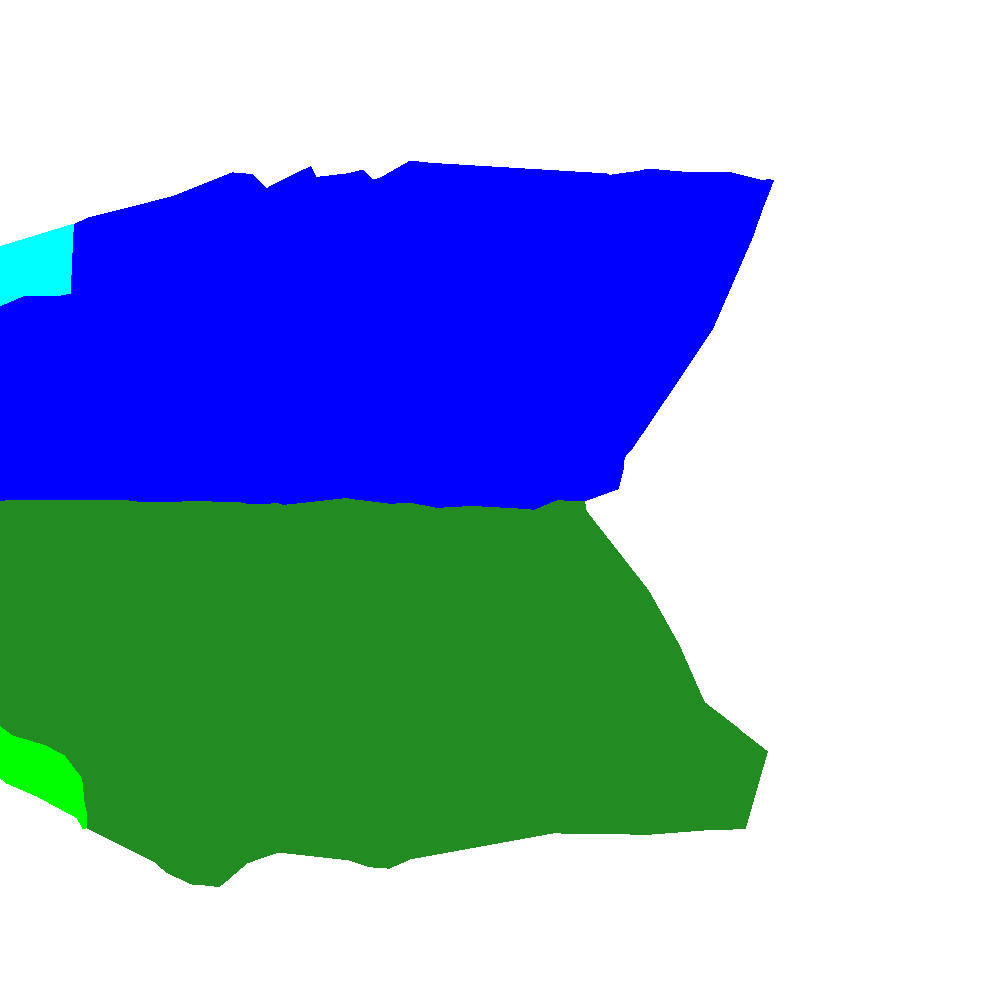};

%\node[aircraft side,fill=black,draw=white,minimum width=1cm,rotate=0,scale = 0.8] at (axis cs:3.0, 0.0) {};
%\node[aircraft side,fill=red,draw=black,minimum width=1cm,rotate=0,scale = 0.8,xscale=-1] at (axis cs:35.0, 0.850) {};

\end{groupplot}

\end{tikzpicture}
	\caption{VerticalCAS policy for the score table (left) and neural network (right) with $\dot{h}_\text{own}=\dot{h}_\text{int} = \SI{0}{\feet\per\minute}$, and $s_{\text{adv}}=\text{COC}$}
	\label{fig:VertCAS_policy}
\end{figure}

\Cref{fig:VertCAS_policy} compares the policies of the VerticalCAS table and neural network representations where each pixel represents the system advisory if the intruder aircraft were at that location. The table uses nearest-neighbor interpolation while the neural network is a function that can be evaluated at any point in the state space. The neural network representation is similar to the discrete table, but there are small differences that may or may not effect performance. \Cref{sec:Verif} describes a formal verification approach to provide safety guarantees when using a neural network collision avoidance system. 

%%%%%%%%%%%%%%%%%
% HorizontalCAS %
%%%%%%%%%%%%%%%%%
\section{HorizontalCAS}\label{sec:HorizontalCAS}
%Another type of aircraft collision avoidance systems issues horizontal turning advisories rather than vertical climb or descend advisories. The system presented here, called HorizontalCAS, is inspired by the ACAS Xu collision avoidance system for unmanned aircraft, which issues horizontal advisories and was tested on a NASA Ikhana unmanned aircraft in 2014~\cite{ACAS-XuTests}. Like ACAS Xa, ACAS Xu uses dynamic programming to compute a lookup table representing state-action values used for decision making, and previous work demonstrates how a neural network approximation could be used to effectively compress the ACAS Xu table~\cite{julian2018deep}. However, ACAS Xu is not publicly available, so this work defines and studies the notional system HorizontalCAS inspired by ACAS Xu, though the safety verification methods presented in \Cref{sec:Verif} could also be applied to ACAS Xu. Supporting code for HorizontalCAS can be found at \url{https://github.com/sisl/HorizontalCAS}.\todo[inline]{KJ: Still need to clean and publish this repository}

\begin{table}
	\centering
	\caption{HorizontalCAS state variables\label{tab:HorStates}}
	\begin{tabular}{lccc}  
	\toprule
	Variable  & Description & Values & Num\\
	\midrule
	$\rho$ (\si{\feet}) & Range to intruder & $[0,50000]$ & 32 \\
	$\theta$ (deg) & Bearing angle to intruder & $[-180,180]$ & 41 \\
	$\psi$ (deg) & Relative heading angle of int. & $[-180,180]$ & 41\\
	$v_\text{own}$ (\si{\feet\per\second}) & Ownship speed & $200$ & 1\\
	$v_\text{int}$  (\si{\feet\per\second}) & Intruder speed & $185$ & 1\\
	$\tau$ (\si{\second}) & Time to loss of vert. separation & $[0,80]$ & 81 \\
	$s_\text{adv}$ & Previous advisory & See \Cref{tab:HorAdv} & 5\\
	\bottomrule
\end{tabular}
\end{table}

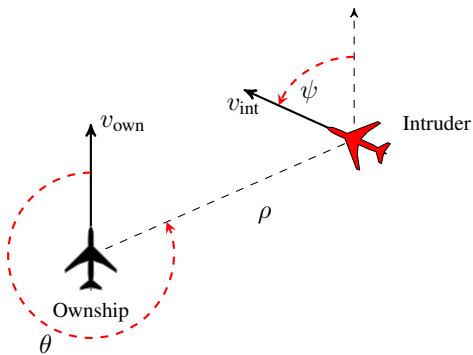
\begin{figure}
	\centering
	%Experimenting
	
\def\layersep{1.6cm}
\def\layerseP{1.12cm}

\begin{tikzpicture}[scale=0.7]

\node[aircraft top,fill=black,draw=white,minimum width=1cm,rotate=90,scale = 0.85] (Own) at (0,0) {} node [below,yshift=-0.5cm,font=\footnotesize] {Ownship};
\coordinate[label=right:$v_\text{own}$] (S0) at (0,2.5);

%\coordinate[label=right:£Intruder£] (Int) at (5,5);
\node[aircraft top,fill=black,draw=white,minimum width=1cm,rotate=156, scale = 0.85] (Int) at (5,2.2) {};
\node at (5,2.2) [above,xshift=1.1cm,font=\footnotesize] {Intruder};
\coordinate[] (IntN) at (5,4.7);
\coordinate[label=below:$v_\text{int}$] (S1) at (2.9,3.15);

\coordinate[label=below:$\rho$] (R) at (3.3,1.1);

% arrows
\draw [thick, ->] (Own) -- (S0);
\draw [dashed,->] (Int) -- (IntN);
\draw [thick, ->] (Int) -- (S1);
            
%draw lines
\draw[dashed] (Own) -- (Int);

%angles
\pic [draw,-stealth,red,thick,dashed,angle radius=1.1cm,"$\psi$"{anchor=west,text = black, below}, angle eccentricity=1] {angle = IntN--Int--S1};
\pic [draw,-stealth,red,thick,dashed,angle radius=1.1cm,"$\theta$"{anchor=west,text = black, below}, angle eccentricity=1] {angle = S0--Own--Int};

\node[aircraft top,fill=black,draw=gray,minimum width=1cm,rotate=90,scale = 0.85]  at (0,0) {};
\node[aircraft top,fill=red,draw=black,minimum width=1cm,rotate=156, scale = 0.85] at (5,2.2) {};

\end{tikzpicture}
	\caption{Aircraft encounter geometry for HorizontalCAS\label{fig:HorizontalCAS}}
\end{figure}

The state space for HorizontalCAS is composed of seven variables that define the encounter, as defined in \Cref{tab:HorStates}. The first five variables define the horizontal geometry and are shown in \Cref{fig:HorizontalCAS}. In this work, ownship and intruder speeds are set to constant values, but the approach can be extended to ranges of speeds as well. Similar to $\tau$ in VerticalCAS, $\tau$ for HorizontalCAS represents a countdown to when the aircraft must be separated horizontally to avoid an NMAC, which occurs when horizontal separation is less than \SI{500}{\feet} and vertical separation is lost. Additionally, $s_\text{adv}$ allows the system to alert consistently.

\begin{table}
	\centering
	\caption{HorizontalCAS advisories}
	\begin{tabular}{lcc}  
	\toprule
	Advisory  & Description & Ownship Turn Rate \\
	\midrule
	COC &  Clear of Conflict & $\left[\SI{-1.5}{\degree\per\second},\ \SI{1.5}{\degree\per\second}\right]$  \\
	WL  &  Weak Left         & $\left[\SI{1.0}{\degree\per\second},\ \SI{2.0}{\degree\per\second}\right]$   \\
	WR  &  Weak Right        & $\left[\SI{-2.0}{\degree\per\second},\ \SI{-1.0}{\degree\per\second}\right]$ \\
	SL  & Strong Left        & $\left[\SI{2.0}{\degree\per\second},\ \SI{4.0}{\degree\per\second}\right]$   \\ 
	SR  & Strong Right       & $\left[\SI{-4.0}{\degree\per\second},\ \SI{-2.0}{\degree\per\second}\right]$ \\ 
	\bottomrule
\end{tabular}
	\label{tab:HorAdv}
\end{table}

The action space consists of five turning advisories of different strengths and directions, which are listed in \Cref{tab:HorAdv}. COC allows the ownship to turn freely while two weak and two strong turning advisories constrain the turn rate. The intruder aircraft is modeled to turn between $\left[ \SI{-1}{\degree\per\second},\ \SI{1}{\degree\per\second} \right]$.

The horizontal dynamics are more complicated than vertical dynamics due to polar coordinates and because the state is defined relative to the ownship's position and heading direction. With ownship and intruder positions of $(x_\text{own},y_\text{own})=(0,0)$ and $(x_\text{int},y_\text{int})=(\rho \cos(\theta),\rho \sin(\theta))$ respectively, and assuming that the ownship and intruder maintain constant turn rates $u_\text{own}$ and $u_\text{int}$ respectively, the aircraft positions after one second will be
\begin{align}
    x'_\text{own} &= \int_0^1 v_\text{own} \cos(u_\text{own}t)dt \\
         &= v_\text{own} \frac{\sin(u_\text{own})}{u_\text{own}} \\
    y'_\text{own} &= \int_0^1 v_\text{own}\sin(u_\text{own}t)dt\\
         &= v_\text{own} \frac{1-\cos(u_\text{own})}{u_\text{own}} \\
    x'_\text{int} &= x_\text{int}+\int_0^1 v_\text{int} \cos(\psi + u_\text{int}t)dt \\
         &= x_\text{int}+v_\text{int} \frac{\sin(\psi+u_\text{int})-\sin(\psi)}{u_\text{int}} \\
    y'_\text{int} &= y_\text{int}+\int_0^1 v_\text{int}\sin(\psi+u_\text{int}t)dt\\
         &= y_\text{int}+v_\text{int} \frac{\cos(\psi)-\cos(\psi+u_\text{int})}{u_\text{int}}.
\end{align}

Once the new positions are computed, the state variables can be updated as 
\begin{equation}\label{eq:HorCAS_dynamics}
    \begin{bmatrix}
\rho \\
\theta \\
\psi \\
v_\text{own} \\
v_\text{int} \\
\tau \\
s_{\text{adv}} \\
\end{bmatrix}
\leftarrow
\begin{bmatrix}
\Vert \left[x'_\text{int}-x'_\text{own},y'_\text{int}-y'_\text{own} \right] \Vert_2 \\
\arctan(y'_\text{int}-y'_\text{own},x'_\text{int}-x'_\text{own}) - u_\text{own} \\
\psi + u_\text{int} - u_\text{own}  \\
v_\text{own} \\
v_\text{int} \\
\max(0,\tau-1) \\
s'_{\text{adv}}
\end{bmatrix}.
\end{equation}
Similar to VerticalCAS, $T(s' \mid s,a)$ is a distribution where the mean turn rate has probability 0.5 and the extremes have probability 0.25.

The reward function is similar to that of VerticalCAS and penalizes NMACs, alerts, reversals, strengthenings, weakenings, and giving COC after a turning advisory when the aircraft are still on a collision course. The HorizontalCAS state space is discretized into 21.8 million states and, like VerticalCAS, solved in a few minutes with Gauss-Seidel value iteration to create a large state-action value table~\cite{Kochenderfer2015chapter4}. This work trains 40 different neural networks, one for each combination of $s_\text{adv}$ and $\tau\in\{0,5,10,15,20,30,40,60\}$. Each neural network has only three inputs, $\rho$, $\theta$, and $\psi$, because the speed dimensions are constant in this work. Each neural network has 5 hidden layers of length 25 and 5 outputs, one for each advisory. When the neural networks are used, the $\tau$ value is rounded down to the closest value in $\tau\in\{0,5,10,15,20,30,40,60\}$ and used with $s_\text{adv}$ to determine the network to be evaluated. Each network was trained for 3000 epochs in 15 minutes each using Tensorflow, resulting in 40 neural networks that reduce required storage from 40.6MB to 442KB while maintaining the correct advisory 97.9\% of the time.

\begin{figure}
	\centering
	\begin{tikzpicture}[]
\begin{groupplot}[height = {5.2cm},width = {8.2cm}, group style={vertical sep=0.35cm, group size=1 by 2}]
\nextgroupplot [ylabel = {Crossrange (kft)},scaled x ticks=false, xticklabels={,,} ,   enlargelimits = false, axis on top]\addplot [point meta min=-3, point meta max=3] graphics [xmin=-6.0, xmax=34.0, ymin=-5.384615384615383, ymax=25.384615384615383]  {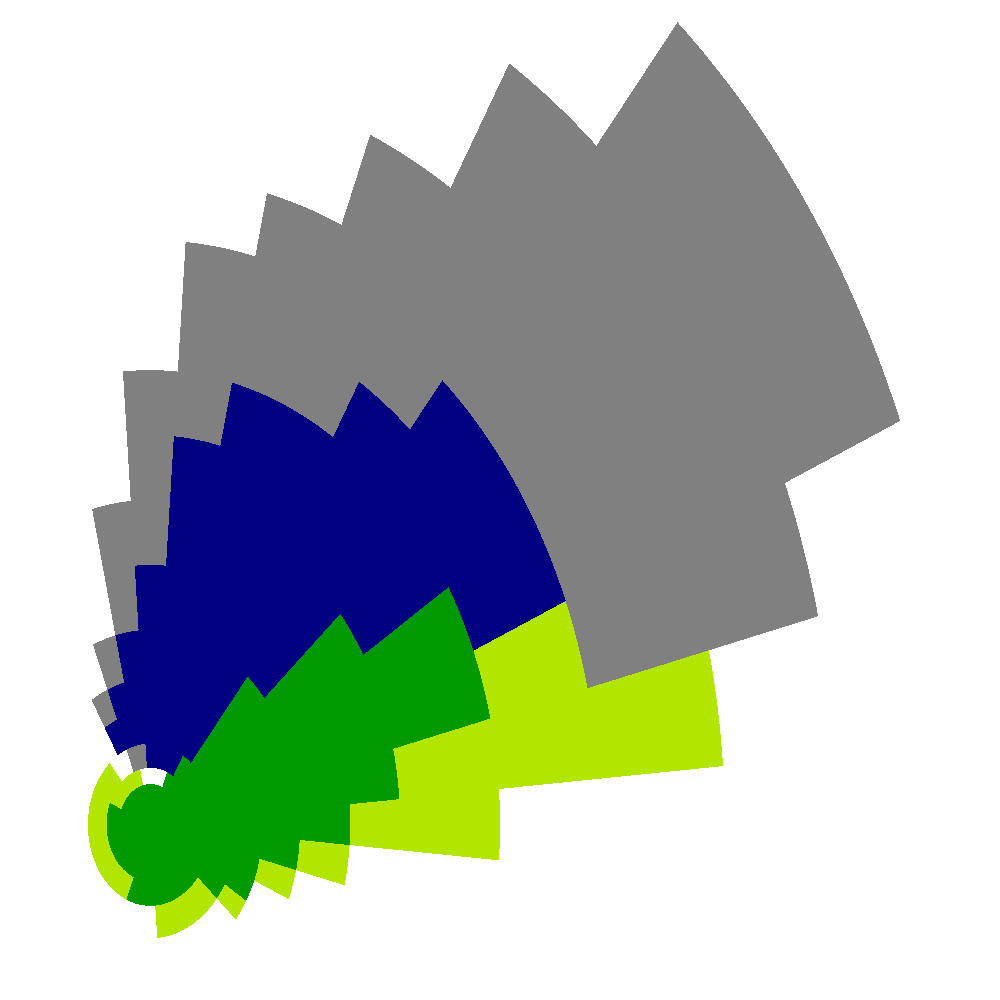};
\node[aircraft top,fill=black,draw=white, minimum width=2.0cm,rotate=0.0,scale = 0.35] at (axis cs:0.0, 0.0) {};;
\node[aircraft top,fill=red,draw=black, minimum width=2.0cm,rotate=-90,scale = 0.35] at (axis cs:30.4, 20.3) {};;

\node[text=black] at (28,-2)  {\footnotesize COC};
\node[text=white] at (6,2)    {\footnotesize SL};
\node[text=white] at (6,9)    {\footnotesize SR};
\node[text=black] at (16,2.6)  {\footnotesize WL};
\node[text=black] at (20,15)  {\footnotesize WR};

\nextgroupplot [ylabel = {Crossrange (kft)}, xlabel = {Downrange (kft)}, enlargelimits = false, axis on top]\addplot [point meta min=-3, point meta max=3] graphics [xmin=-6.0, xmax=34.0, ymin=-5.384615384615383, ymax=25.384615384615383]  {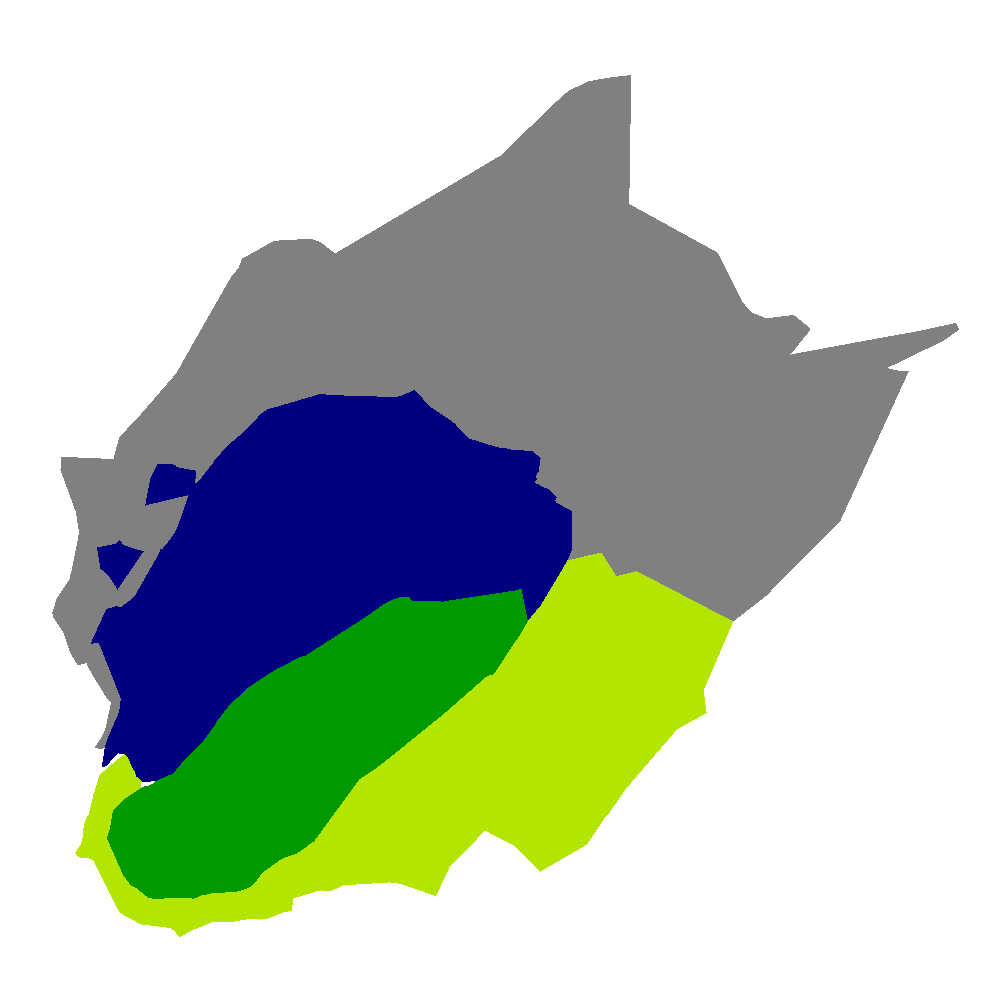};
\node[aircraft top,fill=black,draw=white, minimum width=2.0cm,rotate=0.0,scale = 0.35] at (axis cs:0.0, 0.0) {};;
\node[aircraft top,fill=red,draw=black, minimum width=2.0cm,rotate=-90,scale = 0.35] at (axis cs:30.4, 20.3) {};;

\node[text=black] at (28,-2)  {\footnotesize COC};
\node[text=white] at (6,2)    {\footnotesize SL};
\node[text=white] at (6,9)    {\footnotesize SR};
\node[text=black] at (16,2.6)  {\footnotesize WL};
\node[text=black] at (20,15)  {\footnotesize WR};

\end{groupplot}
\end{tikzpicture}
	\caption{HorizontalCAS policy for the score table (top) and neural network (bottom) with $\psi=\SI{-90}{\degree}$ $\tau=\SI{0}{\second}$, and $s_{\text{adv}}=\text{COC}$}
	\label{fig:HCAS_policy}
\end{figure}
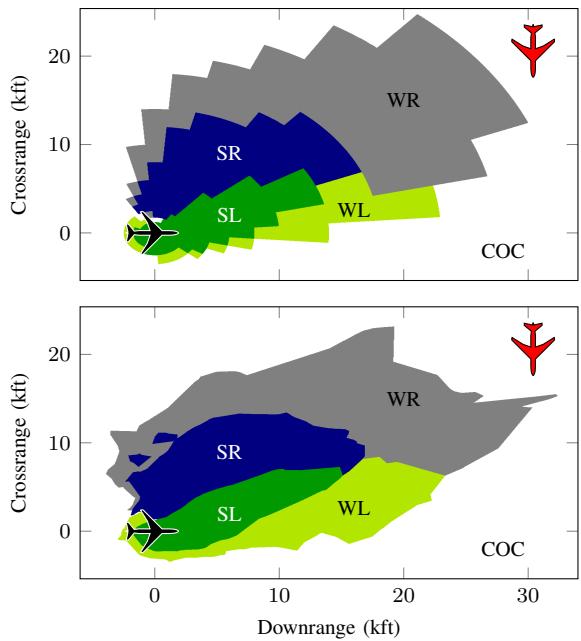

\Cref{fig:HCAS_policy} compares the policies of the HorizontalCAS table and neural network representations. With $\psi$ indicated by the direction of the red aircraft in the upper right corners, each pixel shows the advisory given if the intruder were at that location. The neural network policy is similar to the score table, but there are also key differences. The next section describes a method for guaranteeing safety when the neural network collision avoidance system is used.

%%%%%%%%%%%%%%%%%
% Verification  %
%%%%%%%%%%%%%%%%%
\section{Verification through Reachability}\label{sec:Verif}
%This section presents a general method for verifying closed-loop safety of a neural network controller, while the next two sections discuss how the method can be implemented for vertical and horizontal aircraft collision avoidance neural networks.

The reachability algorithm uses open-source neural network verification tools with the system dynamics to determine which regions of the state space can be reached over time. This method is similar to the work by \citeauthor{xiang2018reachability} but uses discrete actions and incorporates other neural verification tools~\cite{xiang2018reachability}. Neural network verification tools like Reluplex~\cite{katz2017reluplex} and Reluval~\cite{wang2018formal} can determine whether an advisory is given at any point within a region of the input space. These tools can create an over-approximation of the neural networks by splitting the input region into smaller regions, referred to here as cells $c\in\mathcal{C}$, and computing which advisories $\mathcal{A}_c$ can be given within each cell. In the reachability analysis described below, we assume that any of these advisories could be issued from any point within the cell, resulting in an over-approximation of the neural network system. The results presented here use Reluval, which performs quickly for small regions of input space due to its use of symbolic bound propagation~\cite{wang2018formal}.

The analysis begins by initializing a set of reachable cells, $\mathcal{R}_0$, which is the set of states that could occur before the neural network takes action. For collision avoidance networks, this would include all cells where either vertical or horizontal separation is at maximum sensing range. Next, the system dynamics are used to compute $R_{c,a}$, the region of the state space that could be realized at the next time step from some state within $c\in\mathcal{R}_0$ given advisory $a\in\mathcal{A}_c$. If the system dynamics are nonlinear, then $R_{c,a}$ is an over-approximation of the next states reachable from $c$. Then, $\mathcal{R}_1$ is computed as the union of all cells that intersect with $R_{c,a}\ \forall c\in\mathcal{R}_0,\ a\in\mathcal{A}_c$. This process is repeated until either an NMAC cell is added to the reachable set or $\mathcal{R}$ converges to a steady state with no NMAC cells. 

\begin{algorithm}[tb]
	\caption{Reachability analysis for collision avoidance neural networks}
	\label{alg:reach}
	\textbf{Input}: $\mathcal{R}_0$, $\mathcal{C}$
	\begin{algorithmic}[1] %[1] enables line numbers
        \STATE{$t=0$}
		\WHILE{$\text{isSafe}(c)\ \forall c \in \mathcal{R}_t$ and $\mathcal{R}_t \neq \mathcal{R}_{t-1}$}
		\STATE $t=t+1$
		\STATE $\mathcal{R}_{t} \leftarrow \emptyset$
		\FOR{$c \in \mathcal{R}_{t-1}$}
		\STATE Compute $\mathcal{A}_c$ using neural network verification tool
		\FOR{$a \in \mathcal{A}_c$}
		\STATE Compute $R_{c,a}$ using state dynamics
		\FOR{$c' \in \mathcal{C}$}
		\IF{$c' \bigcap R_{c,a} \neq \emptyset$}
		\STATE $\mathcal{R}_t \leftarrow \mathcal{R}_t \bigcup c$
		\ENDIF
		\ENDFOR
		\ENDFOR
		\ENDFOR
		\ENDWHILE
		\IF{$\text{isSafe}(c)\ \forall c \in \mathcal{R}_t$}
		\STATE \textbf{return} Safe
		\ENDIF
		\STATE \textbf{return} Unsafe
	\end{algorithmic}
\end{algorithm}

The reachability analysis is summarized in \Cref{alg:reach}. In general, $\mathcal{R}_t$ may not converge to a steady-state set, but all collision avoidance experiments eventually converged. If $\mathcal{R}_t$ does not converge, then $\mathcal{R}_t$ should be checked for repeating cycles. Otherwise, an inductive argument cannot be used to guarantee safety for all time in the future.

Because the neural networks and system dynamics are over-approximated, the reachability analysis over-approximates the reachable set of real system. As a result, if the over-approximated system cannot reach an unsafe state, then the real neural network system is guaranteed to maintain safety.

%%%%%%%%%%%%%%%%%
% VCAS Results  %
%%%%%%%%%%%%%%%%%
\section{VerticalCAS Results}\label{sec:VertResults}
This section describes the implementation and results of reachability analysis for VerticalCAS neural networks.

\subsection{Implementation}
\begin{figure}
	\centering
	\begin{tikzpicture}[]
\begin{axis}[height = {5.5cm}, width={7.2cm}, xlabel = {$\dot{h}_\text{own}\ (\si{\kilo\feet\per\minute})$}, ylabel = {$h\ (\si{\kilo\feet})$}, enlargelimits = false, axis on top, colormap={mycolormap}{ rgb(0cm)=(0.0,0.0,0.515625) rgb(1cm)=(0.0,0.0,0.53125) rgb(2cm)=(0.0,0.0,0.546875) rgb(3cm)=(0.0,0.0,0.5625) rgb(4cm)=(0.0,0.0,0.578125) rgb(5cm)=(0.0,0.0,0.59375) rgb(6cm)=(0.0,0.0,0.609375) rgb(7cm)=(0.0,0.0,0.625) rgb(8cm)=(0.0,0.0,0.640625) rgb(9cm)=(0.0,0.0,0.65625) rgb(10cm)=(0.0,0.0,0.671875) rgb(11cm)=(0.0,0.0,0.6875) rgb(12cm)=(0.0,0.0,0.703125) rgb(13cm)=(0.0,0.0,0.71875) rgb(14cm)=(0.0,0.0,0.734375) rgb(15cm)=(0.0,0.0,0.75) rgb(16cm)=(0.0,0.0,0.765625) rgb(17cm)=(0.0,0.0,0.78125) rgb(18cm)=(0.0,0.0,0.796875) rgb(19cm)=(0.0,0.0,0.8125) rgb(20cm)=(0.0,0.0,0.828125) rgb(21cm)=(0.0,0.0,0.84375) rgb(22cm)=(0.0,0.0,0.859375) rgb(23cm)=(0.0,0.0,0.875) rgb(24cm)=(0.0,0.0,0.890625) rgb(25cm)=(0.0,0.0,0.90625) rgb(26cm)=(0.0,0.0,0.921875) rgb(27cm)=(0.0,0.0,0.9375) rgb(28cm)=(0.0,0.0,0.953125) rgb(29cm)=(0.0,0.0,0.96875) rgb(30cm)=(0.0,0.0,0.984375) rgb(31cm)=(0.0,0.0,1.0) rgb(32cm)=(0.0,0.015625,1.0) rgb(33cm)=(0.0,0.03125,1.0) rgb(34cm)=(0.0,0.046875,1.0) rgb(35cm)=(0.0,0.0625,1.0) rgb(36cm)=(0.0,0.078125,1.0) rgb(37cm)=(0.0,0.09375,1.0) rgb(38cm)=(0.0,0.109375,1.0) rgb(39cm)=(0.0,0.125,1.0) rgb(40cm)=(0.0,0.140625,1.0) rgb(41cm)=(0.0,0.15625,1.0) rgb(42cm)=(0.0,0.171875,1.0) rgb(43cm)=(0.0,0.1875,1.0) rgb(44cm)=(0.0,0.203125,1.0) rgb(45cm)=(0.0,0.21875,1.0) rgb(46cm)=(0.0,0.234375,1.0) rgb(47cm)=(0.0,0.25,1.0) rgb(48cm)=(0.0,0.265625,1.0) rgb(49cm)=(0.0,0.28125,1.0) rgb(50cm)=(0.0,0.296875,1.0) rgb(51cm)=(0.0,0.3125,1.0) rgb(52cm)=(0.0,0.328125,1.0) rgb(53cm)=(0.0,0.34375,1.0) rgb(54cm)=(0.0,0.359375,1.0) rgb(55cm)=(0.0,0.375,1.0) rgb(56cm)=(0.0,0.390625,1.0) rgb(57cm)=(0.0,0.40625,1.0) rgb(58cm)=(0.0,0.421875,1.0) rgb(59cm)=(0.0,0.4375,1.0) rgb(60cm)=(0.0,0.453125,1.0) rgb(61cm)=(0.0,0.46875,1.0) rgb(62cm)=(0.0,0.484375,1.0) rgb(63cm)=(0.0,0.5,1.0) rgb(64cm)=(0.0,0.515625,1.0) rgb(65cm)=(0.0,0.53125,1.0) rgb(66cm)=(0.0,0.546875,1.0) rgb(67cm)=(0.0,0.5625,1.0) rgb(68cm)=(0.0,0.578125,1.0) rgb(69cm)=(0.0,0.59375,1.0) rgb(70cm)=(0.0,0.609375,1.0) rgb(71cm)=(0.0,0.625,1.0) rgb(72cm)=(0.0,0.640625,1.0) rgb(73cm)=(0.0,0.65625,1.0) rgb(74cm)=(0.0,0.671875,1.0) rgb(75cm)=(0.0,0.6875,1.0) rgb(76cm)=(0.0,0.703125,1.0) rgb(77cm)=(0.0,0.71875,1.0) rgb(78cm)=(0.0,0.734375,1.0) rgb(79cm)=(0.0,0.75,1.0) rgb(80cm)=(0.0,0.765625,1.0) rgb(81cm)=(0.0,0.78125,1.0) rgb(82cm)=(0.0,0.796875,1.0) rgb(83cm)=(0.0,0.8125,1.0) rgb(84cm)=(0.0,0.828125,1.0) rgb(85cm)=(0.0,0.84375,1.0) rgb(86cm)=(0.0,0.859375,1.0) rgb(87cm)=(0.0,0.875,1.0) rgb(88cm)=(0.0,0.890625,1.0) rgb(89cm)=(0.0,0.90625,1.0) rgb(90cm)=(0.0,0.921875,1.0) rgb(91cm)=(0.0,0.9375,1.0) rgb(92cm)=(0.0,0.953125,1.0) rgb(93cm)=(0.0,0.96875,1.0) rgb(94cm)=(0.0,0.984375,1.0) rgb(95cm)=(0.0,1.0,1.0) rgb(96cm)=(0.015625,1.0,0.984375) rgb(97cm)=(0.03125,1.0,0.96875) rgb(98cm)=(0.046875,1.0,0.953125) rgb(99cm)=(0.0625,1.0,0.9375) rgb(100cm)=(0.078125,1.0,0.921875) rgb(101cm)=(0.09375,1.0,0.90625) rgb(102cm)=(0.109375,1.0,0.890625) rgb(103cm)=(0.125,1.0,0.875) rgb(104cm)=(0.140625,1.0,0.859375) rgb(105cm)=(0.15625,1.0,0.84375) rgb(106cm)=(0.171875,1.0,0.828125) rgb(107cm)=(0.1875,1.0,0.8125) rgb(108cm)=(0.203125,1.0,0.796875) rgb(109cm)=(0.21875,1.0,0.78125) rgb(110cm)=(0.234375,1.0,0.765625) rgb(111cm)=(0.25,1.0,0.75) rgb(112cm)=(0.265625,1.0,0.734375) rgb(113cm)=(0.28125,1.0,0.71875) rgb(114cm)=(0.296875,1.0,0.703125) rgb(115cm)=(0.3125,1.0,0.6875) rgb(116cm)=(0.328125,1.0,0.671875) rgb(117cm)=(0.34375,1.0,0.65625) rgb(118cm)=(0.359375,1.0,0.640625) rgb(119cm)=(0.375,1.0,0.625) rgb(120cm)=(0.390625,1.0,0.609375) rgb(121cm)=(0.40625,1.0,0.59375) rgb(122cm)=(0.421875,1.0,0.578125) rgb(123cm)=(0.4375,1.0,0.5625) rgb(124cm)=(0.453125,1.0,0.546875) rgb(125cm)=(0.46875,1.0,0.53125) rgb(126cm)=(0.484375,1.0,0.515625) rgb(127cm)=(0.5,1.0,0.5) rgb(128cm)=(0.515625,1.0,0.484375) rgb(129cm)=(0.53125,1.0,0.46875) rgb(130cm)=(0.546875,1.0,0.453125) rgb(131cm)=(0.5625,1.0,0.4375) rgb(132cm)=(0.578125,1.0,0.421875) rgb(133cm)=(0.59375,1.0,0.40625) rgb(134cm)=(0.609375,1.0,0.390625) rgb(135cm)=(0.625,1.0,0.375) rgb(136cm)=(0.640625,1.0,0.359375) rgb(137cm)=(0.65625,1.0,0.34375) rgb(138cm)=(0.671875,1.0,0.328125) rgb(139cm)=(0.6875,1.0,0.3125) rgb(140cm)=(0.703125,1.0,0.296875) rgb(141cm)=(0.71875,1.0,0.28125) rgb(142cm)=(0.734375,1.0,0.265625) rgb(143cm)=(0.75,1.0,0.25) rgb(144cm)=(0.765625,1.0,0.234375) rgb(145cm)=(0.78125,1.0,0.21875) rgb(146cm)=(0.796875,1.0,0.203125) rgb(147cm)=(0.8125,1.0,0.1875) rgb(148cm)=(0.828125,1.0,0.171875) rgb(149cm)=(0.84375,1.0,0.15625) rgb(150cm)=(0.859375,1.0,0.140625) rgb(151cm)=(0.875,1.0,0.125) rgb(152cm)=(0.890625,1.0,0.109375) rgb(153cm)=(0.90625,1.0,0.09375) rgb(154cm)=(0.921875,1.0,0.078125) rgb(155cm)=(0.9375,1.0,0.0625) rgb(156cm)=(0.953125,1.0,0.046875) rgb(157cm)=(0.96875,1.0,0.03125) rgb(158cm)=(0.984375,1.0,0.015625) rgb(159cm)=(1.0,1.0,0.0) rgb(160cm)=(1.0,0.984375,0.0) rgb(161cm)=(1.0,0.96875,0.0) rgb(162cm)=(1.0,0.953125,0.0) rgb(163cm)=(1.0,0.9375,0.0) rgb(164cm)=(1.0,0.921875,0.0) rgb(165cm)=(1.0,0.90625,0.0) rgb(166cm)=(1.0,0.890625,0.0) rgb(167cm)=(1.0,0.875,0.0) rgb(168cm)=(1.0,0.859375,0.0) rgb(169cm)=(1.0,0.84375,0.0) rgb(170cm)=(1.0,0.828125,0.0) rgb(171cm)=(1.0,0.8125,0.0) rgb(172cm)=(1.0,0.796875,0.0) rgb(173cm)=(1.0,0.78125,0.0) rgb(174cm)=(1.0,0.765625,0.0) rgb(175cm)=(1.0,0.75,0.0) rgb(176cm)=(1.0,0.734375,0.0) rgb(177cm)=(1.0,0.71875,0.0) rgb(178cm)=(1.0,0.703125,0.0) rgb(179cm)=(1.0,0.6875,0.0) rgb(180cm)=(1.0,0.671875,0.0) rgb(181cm)=(1.0,0.65625,0.0) rgb(182cm)=(1.0,0.640625,0.0) rgb(183cm)=(1.0,0.625,0.0) rgb(184cm)=(1.0,0.609375,0.0) rgb(185cm)=(1.0,0.59375,0.0) rgb(186cm)=(1.0,0.578125,0.0) rgb(187cm)=(1.0,0.5625,0.0) rgb(188cm)=(1.0,0.546875,0.0) rgb(189cm)=(1.0,0.53125,0.0) rgb(190cm)=(1.0,0.515625,0.0) rgb(191cm)=(1.0,0.5,0.0) rgb(192cm)=(1.0,0.484375,0.0) rgb(193cm)=(1.0,0.46875,0.0) rgb(194cm)=(1.0,0.453125,0.0) rgb(195cm)=(1.0,0.4375,0.0) rgb(196cm)=(1.0,0.421875,0.0) rgb(197cm)=(1.0,0.40625,0.0) rgb(198cm)=(1.0,0.390625,0.0) rgb(199cm)=(1.0,0.375,0.0) rgb(200cm)=(1.0,0.359375,0.0) rgb(201cm)=(1.0,0.34375,0.0) rgb(202cm)=(1.0,0.328125,0.0) rgb(203cm)=(1.0,0.3125,0.0) rgb(204cm)=(1.0,0.296875,0.0) rgb(205cm)=(1.0,0.28125,0.0) rgb(206cm)=(1.0,0.265625,0.0) rgb(207cm)=(1.0,0.25,0.0) rgb(208cm)=(1.0,0.234375,0.0) rgb(209cm)=(1.0,0.21875,0.0) rgb(210cm)=(1.0,0.203125,0.0) rgb(211cm)=(1.0,0.1875,0.0) rgb(212cm)=(1.0,0.171875,0.0) rgb(213cm)=(1.0,0.15625,0.0) rgb(214cm)=(1.0,0.140625,0.0) rgb(215cm)=(1.0,0.125,0.0) rgb(216cm)=(1.0,0.109375,0.0) rgb(217cm)=(1.0,0.09375,0.0) rgb(218cm)=(1.0,0.078125,0.0) rgb(219cm)=(1.0,0.0625,0.0) rgb(220cm)=(1.0,0.046875,0.0) rgb(221cm)=(1.0,0.03125,0.0) rgb(222cm)=(1.0,0.015625,0.0) rgb(223cm)=(1.0,0.0,0.0) rgb(224cm)=(0.984375,0.0,0.0) rgb(225cm)=(0.96875,0.0,0.0) rgb(226cm)=(0.953125,0.0,0.0) rgb(227cm)=(0.9375,0.0,0.0) rgb(228cm)=(0.921875,0.0,0.0) rgb(229cm)=(0.90625,0.0,0.0) rgb(230cm)=(0.890625,0.0,0.0) rgb(231cm)=(0.875,0.0,0.0) rgb(232cm)=(0.859375,0.0,0.0) rgb(233cm)=(0.84375,0.0,0.0) rgb(234cm)=(0.828125,0.0,0.0) rgb(235cm)=(0.8125,0.0,0.0) rgb(236cm)=(0.796875,0.0,0.0) rgb(237cm)=(0.78125,0.0,0.0) rgb(238cm)=(0.765625,0.0,0.0) rgb(239cm)=(0.75,0.0,0.0) rgb(240cm)=(0.734375,0.0,0.0) rgb(241cm)=(0.71875,0.0,0.0) rgb(242cm)=(0.703125,0.0,0.0) rgb(243cm)=(0.6875,0.0,0.0) rgb(244cm)=(0.671875,0.0,0.0) rgb(245cm)=(0.65625,0.0,0.0) rgb(246cm)=(0.640625,0.0,0.0) rgb(247cm)=(0.625,0.0,0.0) rgb(248cm)=(0.609375,0.0,0.0) rgb(249cm)=(0.59375,0.0,0.0) rgb(250cm)=(0.578125,0.0,0.0) rgb(251cm)=(0.5625,0.0,0.0) rgb(252cm)=(0.546875,0.0,0.0) rgb(253cm)=(0.53125,0.0,0.0) rgb(254cm)=(0.515625,0.0,0.0) }, colorbar, colorbar style = {ylabel={Number of Nearby Cells}}]\addplot [point meta min=0.0, point meta max=800.0] graphics [xmin=-2.5, xmax=2.5, ymin=-3.0, ymax=3.0] {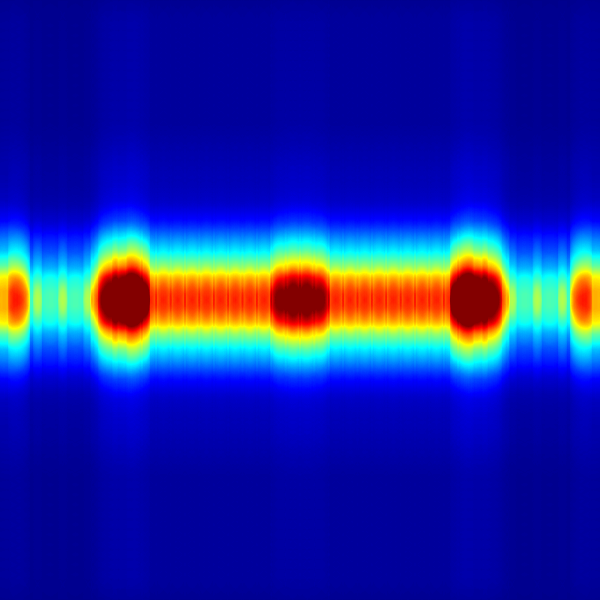};
\end{axis}

\end{tikzpicture}
	\caption{VerticalCAS grid discretization}
	\label{fig:VCAS_density}
\end{figure}

To implement the reachability analysis, the input space must be discretized into cells. Although the networks were created using a discrete table that covered the input space, the table discretization does not need to be used again to define the reachability cells. Smaller cells will decrease over-approximation errors but increase computation time because there will be more cells to cover the input space. Therefore, the input space is split into smaller cells in critical regions of the state space and coarser cells in other locations. For VerticalCAS, regions where intruder separation is low is safety critical, so those cells were made smaller.  
The discretization used here is composed of 3D hyper-rectangles defined by 380 $h$ segments that fill the range between $\SI{-3000}{\feet}$ to $\SI{3000}{\feet}$, 43 $\dot{h}_\text{own}$ segments that fill the range from $\SI{-2500}{\feet\per\min}$ to $\SI{2500}{\feet\per\min}$, and 40 unit-length $\tau$ segments from 0 to 40 for a total of 653600 cells. 
\Cref{fig:VCAS_density} shows a density heat map of the cells, which illustrates that there are more cells close to $h=0$ than where the intruder is far from the ownship.

After defining the cells, Reluval was used to compute all possible advisories given within each cell~\cite{wang2018formal}. All nine advisories where checked with each cell, and all nine neural networks were tested, resulting in 52.5 million queries for Reluval. Reluval required 5.54 CPU-hours to solve all queries and compute $\mathcal{A}_c$ for all cells in all networks, though wall-clock time can be reduced through parallelization since each query is independent.
\begin{table}
	\centering
	\caption{VerticalCAS reachability acceleration limits}
	\begin{tabular}{lcc}  
	\toprule
	Advisory  & $\ddot{h}_{\text{own},\text{min}}$  ($\si{\feet\per\second\squared}$) & $\ddot{h}_{\text{own},\text{max}}$ ($\si{\feet\per\second\squared}$)\\
	\midrule
	COC     &  $-3 - \delta$ & $ 3+\delta$    \\
	DNC or DES1500  &  $-12.2 - \delta$ & $ -12.2+\delta$    \\
	DND  or CL1500  &  $12.2 - \delta$ & $ 12.2+\delta$    \\
	SDES1500 or SDES2500 &  $-13.4 - \delta$ & $ -13.4+\delta$    \\ 
	SCL1500 or SCL2500 &  $13.4 - \delta$ & $ 13.4+\delta$    \\
	
	\bottomrule
\end{tabular}
	\label{tab:VertCAS_ReachLimits}
\end{table}

\begin{figure}
	\centering
	\input{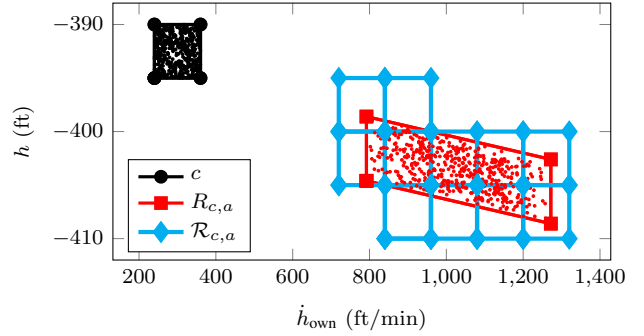}
	\caption{Reachability steps for VerticalCAS}
	\label{fig:VCAS_demo}
\end{figure}

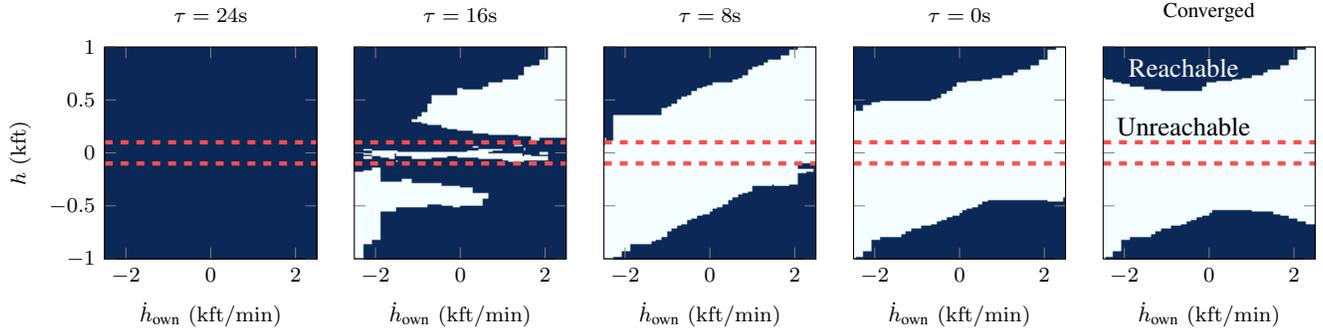
\begin{figure*}
	\centering
	\begin{tikzpicture}[]
\begin{groupplot}[height=4.4cm, width=4.4cm, group style={horizontal sep=0.5cm, group size=5 by 1}]
\nextgroupplot [ylabel = {$h\ (\si{\kilo\feet})$}, title = {$\tau=24\si{\second}$}, xlabel = {$\dot{h}_\text{own}\ (\si{\kilo\feet\per\minute)}$}, enlargelimits = false, axis on top]\addplot [point meta min=0, point meta max=1] graphics [xmin=-2.5, xmax=2.5, ymin=-1.0, ymax=1.0] {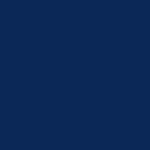};
\addplot+ [red!70!white,dashed,no marks,ultra thick]coordinates {
(-2.5, -0.1)
(2.5, -0.1)
};
\addplot+ [red!70!white,dashed,no marks, ultra thick]coordinates {
(-2.5, 0.1)
(2.5, 0.1)
};
\nextgroupplot [ylabel = {}, title = {$\tau=16\si{\second}$},xlabel = {$\dot{h}_\text{own}\ (\si{\kilo\feet\per\minute)}$}, yticklabels={,,}, scaled y ticks=false, enlargelimits = false, axis on top]\addplot [point meta min=0, point meta max=1] graphics [xmin=-2.5, xmax=2.5, ymin=-1.0, ymax=1.0] {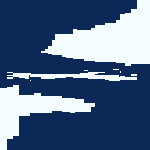};
\addplot+ [red!70!white,dashed,no marks,ultra thick]coordinates {
(-2.5, -0.1)
(2.5, -0.1)
};
\addplot+ [red!70!white,dashed,no marks, ultra thick]coordinates {
(-2.5, 0.1)
(2.5, 0.1)
};
\nextgroupplot [ylabel = {}, title = {$\tau=8\si{\second}$}, xlabel = {$\dot{h}_\text{own}\ (\si{\kilo\feet\per\minute)}$}, yticklabels={,,}, scaled y ticks=false, enlargelimits = false, axis on top]\addplot [point meta min=0, point meta max=1] graphics [xmin=-2.5, xmax=2.5, ymin=-1.0, ymax=1.0] {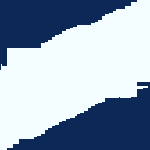};
\addplot+ [red!70!white,dashed,no marks,ultra thick]coordinates {
(-2.5, -0.1)
(2.5, -0.1)
};
\addplot+ [red!70!white,dashed,no marks, ultra thick]coordinates {
(-2.5, 0.1)
(2.5, 0.1)
};
\nextgroupplot [ylabel = {}, title = {$\tau=0\si{\second}$}, xlabel = {$\dot{h}_\text{own}\ (\si{\kilo\feet\per\minute)}$}, yticklabels={,,}, scaled y ticks=false, enlargelimits = false, axis on top]\addplot [point meta min=0, point meta max=1] graphics [xmin=-2.5, xmax=2.5, ymin=-1.0, ymax=1.0] {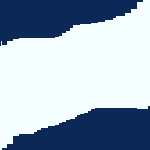};
\addplot+ [red!70!white,dashed,no marks,ultra thick]coordinates {
(-2.5, -0.1)
(2.5, -0.1)
};
\addplot+ [red!70!white,dashed,no marks, ultra thick]coordinates {
(-2.5, 0.1)
(2.5, 0.1)
};
\nextgroupplot [ylabel = {}, title = {Converged}, xlabel = {$\dot{h}_\text{own}\ (\si{\kilo\feet\per\minute)}$}, yticklabels={,,}, scaled y ticks=false, enlargelimits = false, axis on top]\addplot [point meta min=0, point meta max=1] graphics [xmin=-2.5, xmax=2.5, ymin=-1.0, ymax=1.0] {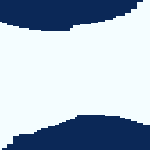};
\addplot+ [red!70!white,dashed,no marks,ultra thick]coordinates {
(-2.5, -0.1)
(2.5, -0.1)
};
\addplot+ [red!70!white,dashed,no marks, ultra thick]coordinates {
(-2.5, 0.1)
(2.5, 0.1)
};
\node at (axis cs:-0.6, 0.8) [white] {Reachable};
\node at (axis cs:-0.6, 0.25) [black] {Unreachable};
\end{groupplot}

\end{tikzpicture}
	\caption{VerticalCAS reachable set over time}
	\label{fig:VC_reachRow}
\end{figure*}

Next, we compute $R_{c,a}$, the region of the state space reachable from cell $c$ given advisory $a\in\mathcal{A}_c$, so a bounded dynamic model is needed. Although the original VerticalCAS MDP uses acceleration constraints for each advisory, reachability analysis would likely not be able to use the same constraints to verify safety. The MDP framework uses a probabilistic model and balances the risk of an NMAC with the cost of alerting; however, reachability analysis is a formal method that determines whether an NMAC is possible with any probability, no matter how small. Therefore, reachability analysis must assume a more constrained dynamic model than what is used in the MDP to verify safety. This section uses $\delta$ to test how relaxed the acceleration bounds can be made before safety cannot be verified. The acceleration bounds are shown in \Cref{tab:VertCAS_ReachLimits}. With $\delta=0$, the non-COC advisory accelerations are constrained to be the average MDP accelerations.

For cell $c$, which defines a small region of the input space where $h \in [h_\text{min},h_\text{max}]$ and $\dot{h}_\text{own} \in [\dot{h}_\text{own,min},\dot{h}_\text{own,max}]$, and advisory $a$, which defines acceleration limits $\ddot{h}_\text{own} \in [\ddot{h}_{\text{own},\text{min}},\ddot{h}_{\text{own},\text{max}}]$, the region of states $(h',\dot{h}'_\text{own})$ reachable at the next time step, $R_{c,a}$, is bounded by
\begin{align}
    h' &\ge h_\text{min} - \dot{h}_\text{own,max} - 0.5 \ddot{h}_{\text{own},\text{max}} \\
    h' &\le h_\text{max} - \dot{h}_\text{own,min} - 0.5 \ddot{h}_{\text{own},\text{min}} \\
    \dot{h}'_\text{own} &\ge \dot{h}_\text{own,min} + \ddot{h}_{\text{own},\text{min}} \\
    \dot{h}'_\text{own} &\le \dot{h}_\text{own,max} + \ddot{h}_{\text{own},\text{max}}\text{.}
\end{align}

Adding $2h'$ to $\dot{h}'_\text{own}$ in \cref{eq:VertCAS_dynamics} eliminates acceleration, leaving $2h'+\dot{h}'_\text{own}=2h-\dot{h}_\text{own}$, which can be bounded to give
\begin{align}
    2h'+\dot{h}'_\text{own} &\ge 2h_\text{min} - \dot{h}_{\text{own},\text{max}} \\
    2h'+\dot{h}'_\text{own} &\le 2h_\text{max} - \dot{h}_{\text{own},\text{min}}\text{.} 
\end{align}

\Cref{fig:VCAS_demo} shows an example cell $c$ with 500 points randomly sampled and propagated forward, which are all contained within $R_{c,a}$. The cyan boxes with diamond corners represent $\mathcal{R}_{c,a}$, the set of cells that overlap with $R_{c,a}$. This process can be repeated for all cells in $c \in \mathcal{R}_t$ with all possible advisories given in the cell to compile $\mathcal{R}_{t+1}$.

The reachability analysis begins with $\mathcal{R}_0$ as the set of all cells with maximum $\tau$, which is the $\tau$ value at which the collision avoidance system begins issuing alerts. $\mathcal{R}_{t+1}$ is computed given $\mathcal{R}_t$ as described above, where $\tau$ counts down by one second with each iteration. In addition, the cells where $h$ is maximum or minimum are added to each $\mathcal{R}_{t+1}$ because an intruder could appear from above or below with less than 40 seconds of horizontal separation. This process continues until there is an NMAC when $\tau=0$, or when $\mathcal{R}_t=\mathcal{R}_{t+1}$.

\subsection{Results}

\Cref{fig:VC_reachRow} shows snapshots of the reachable sets for different $\tau$ values with $\delta=0$ where the dark regions are reachable, the light regions are unreachable, and the dashed red band represents the unsafe NMAC region. Over time, the neural network controller advises the ownship to climb or descend, and by the time $\tau=\SI{0}{\second}$ the reachable set contains no cells within the NMAC region. Furthermore, once $\tau$ reaches 0, the intruder aircraft could be flying in the same direction with the same speed as the ownship, so vertical separation may need to be maintained for all future time to avoid an NMAC. This can be shown by continuing reachable set computation until steady-state is achieved without NMACs 29 seconds after $\tau=0$, so the ownship is guaranteed to be safe for all time.

\begin{table}
	\centering
	\caption{VerticalCAS reachability results for different models}
	\begin{tabular}{cccc}  
	\toprule
	$\delta\ (\si{\feet\per\second\squared})$  & $\tau=0$ min. sep.  $(\si{\feet})$ & Converged min. sep. $(\si{\feet})$ & Safe \\
	\midrule
	0.0 &  390 & 530 &  Yes \\
	1.0 &  375 & 530 &  Yes \\
	2.0 &  215 & 510 &  Yes \\
	3.0 &  200 & 510 &  Yes \\
	4.0 &  113 & 420 &  Yes \\
	5.0 &  101 & 420 &  Yes \\
	6.0 &  12  & N/A &  No  \\
	\bottomrule
\end{tabular}
	\label{tab:VertCAS_Results}
\end{table}

Though the neural network is safe for $\delta=0$, the pilot may respond with different accelerations, so $\delta$ is increased to relax the acceleration limits until an NMAC is reachable. As $\delta$ increases, the pilot can respond more sluggishly, which decreases separation until NMACs become reachable. \Cref{tab:VertCAS_Results} shows that the acceleration bounds can be relaxed by \SI{5.0}{\feet\per\second\squared} while maintaining safety. Reaching an NMAC could be due to over-approximation errors, so the method is unable to prove that the real system will be unsafe.

\subsection{Pilot Delay}
Another component of pilot uncertainty is pilot delay in which the pilot may take a few seconds to respond to an advisory. Reachability can incorporate pilot delay by computing $R_{c,a}$ for any of the previous $\epsilon$ advisories, where $\epsilon$ denotes the number of seconds of pilot delay. The reachability analysis tracks which cells are reachable along with the sequences of $\epsilon$ previous advisories that could be given to reach each cell. 

\begin{figure}
	\centering
	\begin{tikzpicture}[]
\begin{groupplot}[height=4.6cm, width=4.6cm, group style={horizontal sep=0.5cm, group size=2 by 1}]
\nextgroupplot [ylabel = {$h\ (\si{\kilo\feet})$},xlabel = {$\dot{h}_\text{own}\ (\si{\kilo\feet\per\minute})$}, enlargelimits = false, axis on top]\addplot [point meta min=0, point meta max=1] graphics [xmin=-2.5, xmax=2.5, ymin=-1.0, ymax=1.0] {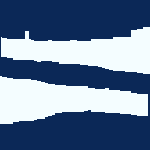};
\addplot+ [red!70!white,dashed,no marks,ultra thick]coordinates {
(-2.5, -0.1)
(2.5, -0.1)
};
\addplot+ [red!70!white,dashed,no marks, ultra thick]coordinates {
(-2.5, 0.1)
(2.5, 0.1)
};

\nextgroupplot [xlabel = {$\dot{h}_\text{own}\ (\si{\kilo\feet\per\minute})$}, yticklabels={,,}, scaled y ticks=false, enlargelimits = false, axis on top]\addplot [point meta min=0, point meta max=1] graphics [xmin=-2.5, xmax=2.5, ymin=-1.0, ymax=1.0] {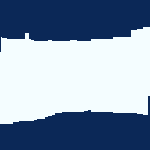};
\addplot+ [red!70!white,dashed,no marks,ultra thick]coordinates {
(-2.5, -0.1)
(2.5, -0.1)
};
\addplot+ [red!70!white,dashed,no marks, ultra thick]coordinates {
(-2.5, 0.1)
(2.5, 0.1)
};
\node at (axis cs:-0.6, 0.8) [white] {Reachable};
\node at (axis cs:-0.6, 0.25) [black] {Unreachable};
\end{groupplot}

\end{tikzpicture}
	\caption{VerticalCAS reachable set at $\tau=0$ for nominal system (left) and system that prevents multiple reversals (right) for three second pilot delay}
	\label{fig:VC_pd}
\end{figure}
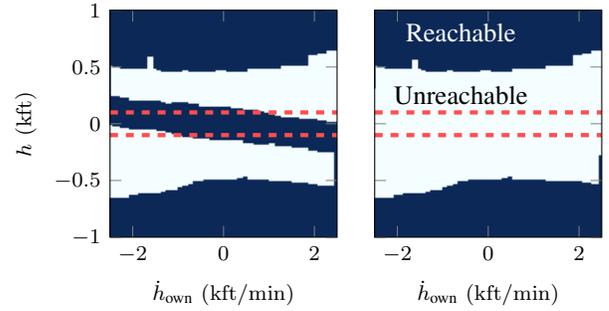

With $\delta=0$ and $\epsilon=3$, there are many reachable NMAC states, as shown in the left plot of \Cref{fig:VC_pd}. These states exist because the neural network continuously reverses the alerting direction after the pilot ignores the advisory. For example, if the intruder is just above the ownship, the system will advise descend. However, if the pilot instead follows an older advisory and climbs instead, the ownship will pass above the intruder where the system may reverse its advisory to climb. 

To address this issue, an online cost that prevents multiple reversals could be used. This online cost can be modeled by tracking the number of reversals along with $\epsilon$ previous advisories and continuing the previous advisory in cases where the system would issue a second reversal. The right plot in \Cref{fig:VC_pd} shows that NMACs will be avoided with pilot delay if online costs are added to prevent multiple reversals.

%%%%%%%%%%%%%%%%%
% HCAS Results  %
%%%%%%%%%%%%%%%%%
\section{HorizontalCAS Reachability}\label{sec:HorResults}
This section describes the implementation and results of reachability analysis for HorizontalCAS neural networks.

\subsection{Implementation}
To implement reachability, bounds on the state variables at the next time step, $R_{c,a}$ must be computed for some bounded input region, $c$. Unlike VerticalCAS, the state dynamics are non-linear and coupled as the polar coordinates are transformed to Cartesian, updated, and returned to polar form. As a result, computing $R_{C,a}$ in polar-coordinates is difficult and results in loose bounds, which significantly increases over-approximation errors. However, reachability with Cartesian coordinates results in less over-approximation error, so this work considers neural networks that use Cartesian coordinates as inputs rather than polar coordinates. The same HorizontalCAS MDP can be used to generate the table with state variables $\rho$ and $\theta$, but the neural network training data uses $x=\rho\cos(\theta)$ and $y=\rho\sin(\theta)$ instead of $\rho$ and $\theta$. 

%\begin{figure}
%	\centering
%	\input{HCAS_Density.tex}
%	\caption{HorizontalCAS grid discretization}
%	\label{fig:HCAS_density}
%\end{figure}

Similar to VerticalCAS, the cells are discretized in $x$ (downrange) and $y$ (crossrange) more heavily near the NMAC region and areas leading to that region.%, as shown in \Cref{fig:HCAS_density}. 
In addition, $\psi$ was discretized to 360 one-degree segments. The final discretization has 74.1 million cells and 14.8 billion Reluval queries, which required 227 CPU-hours to compute $\mathcal{A}_c$ for all cells in all 40 networks.

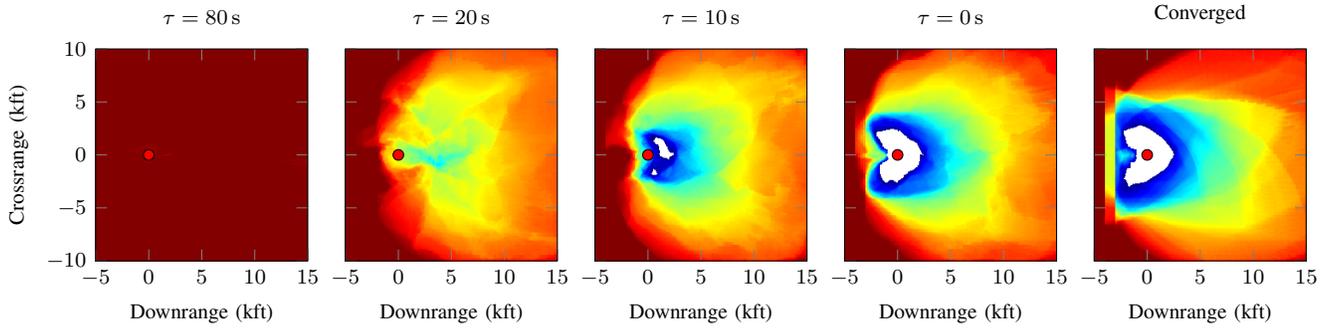
\begin{figure*}
	\centering
	\begin{tikzpicture}[]
\begin{groupplot}[height=4.4cm, width=4.4cm, group style={horizontal sep=0.5cm, group size=5 by 1}]
\nextgroupplot [ylabel = {Crossrange (kft)}, title = {$\tau=\SI{80}{\second}$}, xlabel = {Downrange (kft)}, , enlargelimits = false, axis on top]\addplot [point meta min=0, point meta max=360] graphics [xmin=-5, xmax=15, ymin=-10, ymax=10] {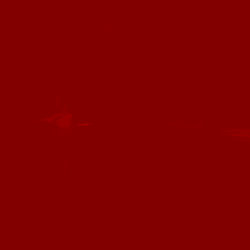};
\draw[black,fill=red] (axis cs:0, 0) circle[radius=0.5];
\nextgroupplot [ylabel = {}, title = {$\tau=\SI{20}{\second}$}, xlabel = {Downrange (kft)}, yticklabels={,,}, scaled y ticks=false, enlargelimits = false, axis on top]\addplot [point meta min=0, point meta max=360] graphics [xmin=-5, xmax=15, ymin=-10, ymax=10] {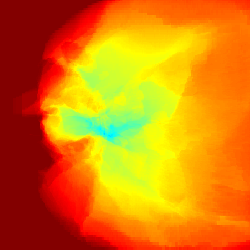};
\draw[black,fill=red] (axis cs:0, 0) circle[radius=0.5];
\nextgroupplot [ylabel = {}, title = {$\tau=\SI{10}{\second}$}, xlabel = {Downrange (kft)}, yticklabels={,,}, scaled y ticks=false, enlargelimits = false, axis on top]\addplot [point meta min=0, point meta max=360] graphics [xmin=-5, xmax=15, ymin=-10, ymax=10] {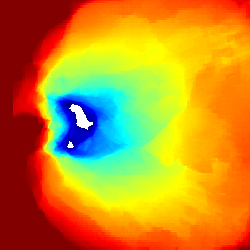};
\draw[black,fill=red] (axis cs:0, 0) circle[radius=0.5];
\nextgroupplot [ylabel = {}, title = {$\tau=\SI{0}{\second}$}, xlabel = {Downrange (kft)}, yticklabels={,,}, scaled y ticks=false, enlargelimits = false, axis on top]\addplot [point meta min=0, point meta max=360] graphics [xmin=-5, xmax=15, ymin=-10, ymax=10] {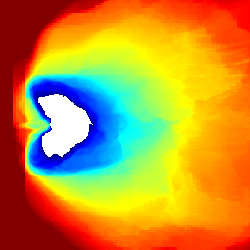};
\draw[black,fill=red] (axis cs:0, 0) circle[radius=0.5];
\nextgroupplot [ylabel = {}, title = {Converged}, xlabel = {Downrange (kft)}, yticklabels={,,}, scaled y ticks=false, enlargelimits = false, axis on top]\addplot [point meta min=0, point meta max=360] graphics [xmin=-5, xmax=15, ymin=-10, ymax=10] {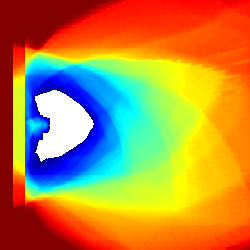};
\draw[black,fill=red] (axis cs:0, 0) circle[radius=0.5];
\end{groupplot}

\end{tikzpicture}
	\caption{HorizontalCAS reachable sets over time}
	\label{fig:HC_reachRow}
\end{figure*}

\begin{table}
	\centering
	\caption{HorizontalCAS reachability turn rate limits}
	\begin{tabular}{lccc}  
	\toprule
	Aircraft & Advisory  & $u_\text{min}$  ($\si{\degree\per\second}$) & $u_\text{max}$ ($\si{\degree\per\second}$)\\
	\midrule
    Ownship & COC     &  $- \delta$ & $ \delta$    \\
	Ownship & WL &  $1.5- \delta$ & $ 1.5+\delta$    \\
	Ownship & WR &  $-1.5 - \delta$ & $ -1.5+\delta$    \\
	Ownship & SR &  $3.5 - \delta$ & $ 3.5+\delta$    \\ 
	Ownship & SL &  $-3.5 - \delta$ & $ -3.5+\delta$    \\
	Intruder & N/A & $-\delta$ & $\delta$ \\
	\bottomrule
\end{tabular}
	\label{tab:HCAS_ReachLimits}
\end{table}

Next, reachability analysis must assume a constrained dynamic model. As with VerticalCAS, the variable $\delta$ is used to relax the turn rate bounds as shown in \Cref{tab:VertCAS_ReachLimits}.

Computing $R_{c,a}$ follows the same method as for VerticalCAS where upper and lower bounds are computed for each state variable. The terms in \Cref{eq:HorCAS_dynamics} can be rearranged using Ptolemy's trigonometric identities as
\begin{align}
    x' &= x + v_\text{int}\cos(\psi)-v_\text{own} - v_\text{own}e_1(u_0) \\
    &\quad-v_\text{int}\sin(\psi)e_2(u_1)-v_1\cos(\psi)e_1(u_1) \\
    y' &= y + v_\text{int}\sin(\psi) - v_\text{own}e_2(u_\text{own}) \\
    &\quad+v_\text{int}\cos(\psi)e_2(u_1)-v_\text{int}\sin(\psi)e_1(u_1)\text{,}
\end{align}
where functions $e_1$ and $e_2$ are 
\begin{align}
   e_1(x) &=  \frac{x-\sin(x)}{x} \\
   e_2(x) &= \frac{1-\cos(x)}{x}.
\end{align}

Functions $e_1$ and $e_2$ represent error functions due to turning that evaluate to zero when the input is zero. For small bounded turning rates, we can compute bounds on $e_1$ and $e_2$ as well as the other terms in the expressions for $x'$ and $y'$. To compute bounds on $x'$ and $y'$, each term in the expressions can be bounded individually in order to decouple the terms, and then bounds on $x'$ and $y'$ are computed considering the minimum or maximum of each term in the summation. Lastly, the bounds on $x'$ and $y'$ must be rotated because the coordinate frame of $x$ and $y$ is fixed to the ownship. Again, simple upper and lower bounds are computed for a bounded ownship turn rate, leading to the final bounds on intruder location at the next time step. Bounds on $\psi'$ are easily computed as 
\begin{align}
    \psi' &\ge \psi_\text{min} + u_\text{int,min} - u_\text{own,max} \\
    \psi' &\le \psi_\text{max} + u_\text{int,max} - u_\text{own,min}.
\end{align}
The resulting $R_{c,a}$ is a hyper-rectangle that over-approximates the set of states reachable from any state within $c$. 

Similar to VerticalCAS, $\mathcal{R}_0$ is initialized as all cells with $\tau=\SI{80}{\second}$, the maximum value. The reachable set of cells at the next time step is computed iteratively, and the cells on the boundary of the sensing region are always added to the reachable set to simulate an intruder appearing closer in altitude but far away horizontally. 

\subsection{Results} 
\Cref{fig:HC_reachRow} shows snapshots of the reachable cells at different $\tau$ values. Because the cells are three-dimensional but the plots are two-dimensional, the color of the cell indicates the number of cells reachable for that $(x,y)$ location, i.e. the number of intruder heading angles reachable at that location. The darker red indicates more cells are reachable while blue indicates fewer cells are reachable, and white indicates no cells are reachable. The red dot at the origin represents the NMAC region. Initially, all cells are added to the reach state at $\tau=\SI{80}{\second}$, but over time the neural network collision avoidance system clears an unreachable around the NMAC region. Because the NMAC region is completely inside the white unreachable region when $\tau=\SI{0}{\second}$, no NMACs are possible at that time. Furthermore, each iteration beyond $\tau=\SI{0}{\second}$ maintains a safe unreachable region around the NMAC region, and after 115 seconds beyond $\tau=\SI{0}{\second}$, the reachable set converges to a steady-state set. As a result, HorizontalCAS is guaranteed to avoid NMACs for the bounded dynamic model with $\delta=0$.

\begin{figure}
	\centering
	\begin{tikzpicture}[]
\begin{groupplot}[height=4.6cm, width=4.6cm, group style={horizontal sep=0.5cm, group size=2 by 1}]
\nextgroupplot [ylabel = {Crossrange (kft)}, xlabel = {Downrange (kft)}, enlargelimits = false, axis on top]
\addplot [point meta min=0, point meta max=360] graphics [xmin=-5, xmax=15, ymin=-10, ymax=10] {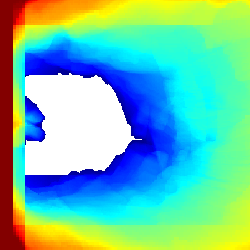};
\draw[black,fill=red] (axis cs:0, 0) circle[radius=0.5];

\nextgroupplot [yticklabels={,,}, scaled y ticks=false,xlabel = {Downrange (kft)}, enlargelimits = false, axis on top]
\addplot [point meta min=0, point meta max=360] graphics [xmin=-5, xmax=15, ymin=-10, ymax=10] {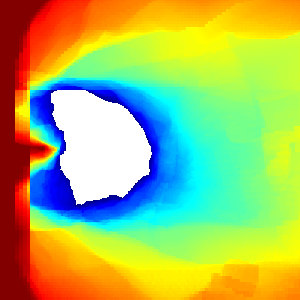};
\draw[black,fill=red] (axis cs:0, 0) circle[radius=0.5];

\end{groupplot}

\end{tikzpicture}
	\caption{HorizontalCAS reachable set for the retrained networks with $\delta=0.0$ (left) and $\delta=0.2$ (right) once $\tau$ reaches 0}
	\label{fig:HCAS_retrained}
\end{figure}

The reachability method proves separation of $\SI{1046}{\feet}$ with the current neural networks, but increasing $\delta$ slightly reduces the safety margin and results in an NMAC. To prove safety for more relaxed dynamics, either over-approximation errors must be reduced by, for example, decreasing cell sizes, or a more robust neural network must be trained. We created a more robust neural network collision avoidance system by reducing the alerting cost when generating the MDP cost table, creating larger alerting regions. As a result, the neural network better maintains separation but alerts more frequently. \Cref{fig:HCAS_retrained} shows the final reachable sets for the retrained networks for two $\delta$ values, which shows that the new neural networks can avoid NMACs for $\delta=0.2$.

Additionally, initial studies with $v_\text{int}=v_\text{own}$ could not verify safety. Over-approximation errors allowed the intruder aircraft to move closer to the ownship than what would really be possible, which eventually leads to reaching an NMAC. In general, this reachability method cannot guarantee safety in cases where the intruder aircraft has as much or more control authority than the ownship because the intruder will be able to chase the ownship into an NMAC.

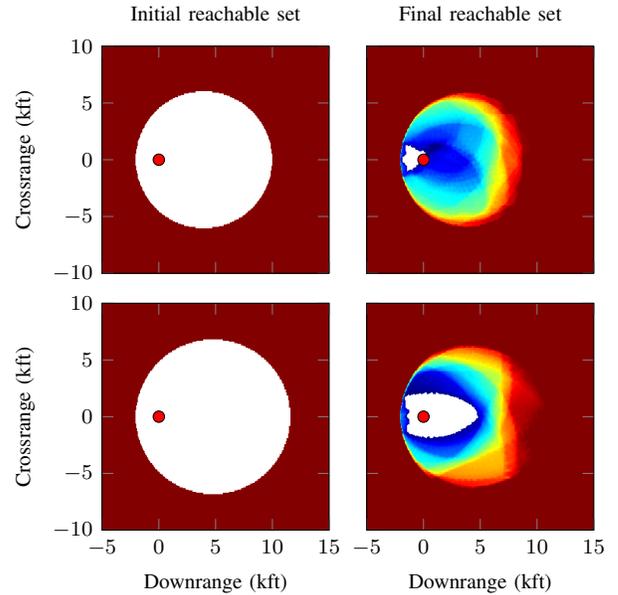
\begin{figure}
	\centering
	\begin{tikzpicture}[]
\begin{groupplot}[height=4.6cm, width=4.6cm, group style={vertical sep=0.4cm, horizontal sep=0.5cm, group size=2 by 2}]

\nextgroupplot [title={Initial reachable set}, ylabel = {Crossrange (kft)}, scaled x ticks=false, xticklabels={,,} , enlargelimits = false, axis on top]\addplot [point meta min=0, point meta max=360] graphics [xmin=-5, xmax=15, ymin=-10, ymax=10] {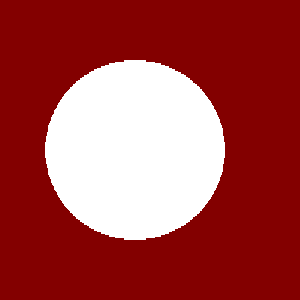};
\draw[black,fill=red] (axis cs:0, 0) circle[radius=0.5];
\nextgroupplot [title={Final reachable set}, ylabel = {}, scaled x ticks=false, xticklabels={,,}, yticklabels={,,}, scaled y ticks=false, enlargelimits = false, axis on top]\addplot [point meta min=0, point meta max=360] graphics [xmin=-5, xmax=15, ymin=-10, ymax=10] {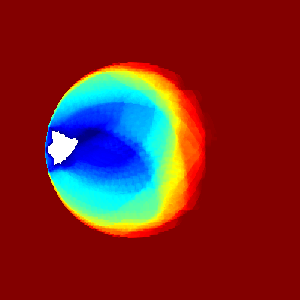};
\draw[black,fill=red] (axis cs:0, 0) circle[radius=0.5];

\nextgroupplot [ylabel = {Crossrange (kft)}, xlabel = {Downrange (kft)}, enlargelimits = false, axis on top]\addplot [point meta min=0, point meta max=360] graphics [xmin=-5, xmax=15, ymin=-10, ymax=10] {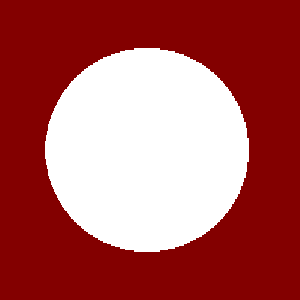};
\draw[black,fill=red] (axis cs:0, 0) circle[radius=0.5];
\nextgroupplot [ylabel = {}, xlabel = {Downrange (kft)},yticklabels={,,},  scaled y ticks=false, enlargelimits = false, axis on top]\addplot [point meta min=0, point meta max=360] graphics [xmin=-5, xmax=15, ymin=-10, ymax=10] {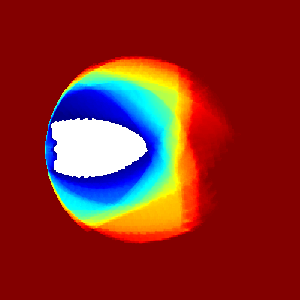};
\draw[black,fill=red] (axis cs:0, 0) circle[radius=0.5];

\end{groupplot}

\end{tikzpicture}
	\caption{Verifying which initial reachable sets can be made safe for an unsafe set (top row) and safe set (bottom row)}
	\label{fig:HCAS_circle}
\end{figure}

\subsection{Safe Regions}
Although HorizontalCAS begins alerting whenever an intruder enters the boundary of its sensing region, situations could arise where an intruder is ignored until it is closer to the ownship. For example, if there are multiple intruders present, the ownship may be forced to get closer to one of the intruders while avoiding the other. Reachability analysis can be used to compute the largest region of intruder locations where an NMAC will still be avoided. 

The left plots of \Cref{fig:HCAS_circle} show sets for cells that satisfy
\begin{equation}
    (x-v_\text{own}t)^2 + y^2 \ge (v_\text{own}t+d)^2\text{,}
\end{equation}
where $t$ is the amount of time for the ownship to maintain separation of at least $d$ feet from the intruder if the ownship flies straight. With $d=\SI{2000}{\feet}$, the top row of \Cref{fig:HCAS_circle} uses $t=\SI{20}{\second}$ while the bottom row uses $t=\SI{24}{\second}$. The longer time results in a larger circle omitted from the initial reachable set, which is the difference between an NMAC and guaranteed safety. This result signifies that the entire state space does not need to be analyzed to prove safety, which could lead to methods that improve computation speed by relaxing cell discretization in areas of the state space that are unimportant.

\subsection{Sensor Error}
The reachability method can model sensor error by expanding the cells' boundaries when computing $\mathcal{A}_\text{c}$ by some amount of bounded sensor measurement error. The cell bounds on $x$ and $y$ were expanded outward by 5\% of the range to the cell center to model position error estimates of up to 5\%. However, making the cells larger increases the computation time, and Reluval required two orders of magnitude longer to compute $\mathcal{A}_{c}$ than without sensor error. In addition, safety could not be verified because many cells issue both SL and SR advisories from all possible $s_\text{adv}$, which means the network could continuously issue reversals until reaching an NMAC. This issue can be fixed by more heavily penalizing reversals or adding online costs to prevent multiple reversals.

\section{Conclusions}\label{sec:conc}
We presented two open-source frameworks for generating collision avoidance tables and neural networks inspired by prototypes of the ACAS X systems. The VerticalCAS system issues vertical rate advisories to avoid intruder aircraft while the HorizontalCAS system issues turning rate advisories. A reachability method was introduced that uses existing neural network verification tools to over-approximate the neural network and computes all ways the input space will evolve in the closed-loop system. The method was used for both HorizontalCAS and VerticalCAS neural network systems and demonstrates that both systems can be proven safe subject to bounded state dynamics and pilot models. Trade-offs between the looseness of dynamic bounds and guaranteed safety margins were explored. Pilot delay was incorporated, and issues stemming from delay were addressed with online costs. Furthermore, we demonstrated a method to generate a large initial reachable sets that can still be proven safe, which can be used to determine the region where the neural networks decisions are critical for safety. Future work will study ways to autonomously and efficiently discretize the input region to reduce over-approximation errors in safety-critical regions without adding more cells than necessary. The closed-loop reachability approach suggests a method to one day verify neural network controllers for use in safety critical systems. 

\section*{Acknowledgments}
This material is based upon work supported by the National Science Foundation Graduate Research Fellowship under Grant No. DGE$-$1656518. Any opinion, findings, and conclusions or recommendations expressed in this material are those of the authors and do not necessarily reflect the views of the National Science Foundation.

%\clearpage

\printbibliography

@String { dasc        = {Digital Avionics Systems Conference (DASC)} }

@String { icassp      = {International Conference on Acoustics, Speech, and Signal Processing (ICASSP)} }

@String { jgcd        = {AIAA Journal on Guidance, Control, and Dynamics} }

@String { mit         = {Massachusetts Institute of Technology} }

@article{bellman1952theory,
	title={On the theory of dynamic programming},
	author={Bellman, Richard},
	journal={Proceedings of the National Academy of Sciences of the United States of America},
	volume={38},
	number={8},
	pages={716},
	year={1952},
}

@InCollection{Kochenderfer2015chapter4,
	Title                    = {Sequential Problems},
	Author                   = {Mykel J. Kochenderfer},
	Booktitle                = {Decision Making under Uncertainty: Theory and Application},
	Publisher                = {MIT Press},
	Year                     = {2015},
	chapter     = {4},
	pages = {77--112},
	
}

@TechReport{kochenderfer2011robust,
  Title                    = {Robust Airborne Collision Avoidance through Dynamic Programming},
  Author                   = {Mykel J. Kochenderfer and James P. Chryssanthacopoulos},
  Institution              = {Massachusetts Institute of Technology, Lincoln Laboratory},
  Year                     = {2011},
  Number                   = {ATC-371},
  Type                     = {Project Report},
}

@article{julian2018deep,
  title={Deep neural network compression for aircraft collision avoidance systems},
  author={Julian, Kyle D and Kochenderfer, Mykel J and Owen, Michael P},
  journal=jgcd,
  volume={42},
  number={3},
  pages={598--608},
  year={2018},
  publisher={American Institute of Aeronautics and Astronautics}
}

@inproceedings{julian2016policy,
  title={Policy compression for aircraft collision avoidance systems},
  author={Julian, Kyle D and Lopez, Jessica and Brush, Jeffrey S and Owen, Michael P and Kochenderfer, Mykel J},
  booktitle=dasc,
  year={2016},
  organization={IEEE}
}

@article{julian2019verifying,
  title={Verifying aircraft collision avoidance neural networks through linear approximations of safe regions},
  author={Julian, Kyle D and Sharma, Shivam and Jeannin, Jean-Baptiste and Kochenderfer, Mykel J},
  journal={arXiv preprint arXiv:1903.00762},
  year={2019}
}

@article{julian2019reachability,
  title={A Reachability Method for Verifying Dynamical Systems with Deep Neural Network Controllers},
  author={Julian, Kyle D and Kochenderfer, Mykel J},
  journal={arXiv preprint arXiv:1903.00520},
  year={2019}
}

@article{xiang2018reachability,
	title={Reachability analysis and safety verification for neural network control systems},
	author={Xiang, Weiming and Johnson, Taylor T},
	journal={arXiv preprint arXiv:1805.09944},
	year={2018}
}

@InProceedings{ReLU,
  author    = {G. E. Dahl and T. N. Sainath and G. E. Hinton},
  title     = {Improving deep neural networks for LVCSR using rectified linear units and dropout},
  booktitle = icassp,
  year      = {2013},
  pages     = {8609-8613},
  doi       = {10.1109/ICASSP.2013.6639346},
  issn      = {1520-6149},
}

@inproceedings{akintunde2018reachability,
	title={Reachability Analysis for Neural Agent-Environment Systems},
	author={Akintunde, Michael and Lomuscio, Alessio and Maganti, Lalit and Pirovano, Edoardo},
	booktitle={International Conference on Principles of Knowledge Representation and Reasoning},
	year={2018}
}

@article{ivanov2018verisig,
	title={Verisig: verifying safety properties of hybrid systems with neural network controllers},
	author={Ivanov, Radoslav and Weimer, James and Alur, Rajeev and Pappas, George J and Lee, Insup},
	journal={arXiv preprint arXiv:1811.01828},
	year={2018}
}

@InCollection{Kochenderfer2015chapter10,
	Title                    = {Optimized Airborne Collision Avoidance},
	Author                   = {Mykel J. Kochenderfer},
	Booktitle                = {Decision Making under Uncertainty: Theory and Application},
	Publisher                = {MIT Press},
	Year                     = {2015},
	chapter     = {10},
	pages = {249--273},
	
}

@article{liu2019algorithms,
	author       = {Liu, Changliu and Arnon, Tomer and Lazarus, Christopher and Barrett, Clark and Kochenderfer, Mykel J},
	title        = {Algorithms for Verifying Deep Neural Networks},
	journal      = {arXiv preprint arXiv:1903.06758},
	year         = {2019},	
}

@TechReport{ACAS-XuTests,
  Title                    = {{ACAS}-{X}u Initial Self-Separation Flight Tests},
  Author                   = {Marston, Mike and Baca, Gabe},
  Institution              = {NASA},
  Year                     = {2015},
  Number                   = {DFRC-E-DAA-TN22968},

}

@inproceedings{wang2018formal,
	title={Formal security analysis of neural networks using symbolic intervals},
	author={Wang, Shiqi and Pei, Kexin and Whitehouse, Justin and Yang, Junfeng and Jana, Suman},
	booktitle={Security Symposium ($\{$USENIX$\}$ Security 18)},
	pages={1599--1614},
	year={2018}
}

@InProceedings{katz2017reluplex,
	author       = {Katz, Guy and Barrett, Clark and Dill, David L and Julian, Kyle D and Kochenderfer, Mykel J},
	title        = {Reluplex: An efficient {SMT} solver for verifying deep neural networks},
	booktitle    = {International Conference on Computer Aided Verification},
	year         = {2017},
	pages        = {97--117},
	organization = {Springer},
	doi          = {10.1007/978-3-319-63387-9\_5},
}
\end{document}